\documentclass[a4paper,fleqn]{cas-dc}

\usepackage[authoryear]{natbib}
\usepackage{amsmath}

\usepackage[outdir=./]{epstopdf}
\usepackage{epsfig,epsf,rotating}
\usepackage{subfigure}
\usepackage{epsf}
\usepackage{placeins}
\usepackage{hyperref}

\def\tsc#1{\csdef{#1}{\textsc{\lowercase{#1}}\xspace}}
\tsc{WGM}
\tsc{QE}

\begin{document}
\let\WriteBookmarks\relax
\def\floatpagepagefraction{1}
\def\textpagefraction{.001}

\shorttitle{Application of Machine Learning in Seismic Fragility Assessment of Bridges with SRR Columns}    

\shortauthors{Akbarnezhad et al.}  

\title [mode = title]{Application of Machine Learning in Seismic Fragility Assessment of Bridges with SMA-Restrained Rocking Columns}  



%

\author[1]{Miles Akbarnezhad}[orcid=0000-0002-2771-6215]
\author[1]{Mohammad Salehi}[orcid=0000-0003-1715-6282]
\cormark[1]
\author[1]{Reginald DesRoches}





\affiliation[1]{organization={Department of Civil and Environmental Engineering, Rice University},
            addressline={6100 Main St}, 
            city={Houston},
            postcode={77005}, 
            state={TX},
            country={USA}
}

\begin{abstract}
This paper evaluates the seismic fragility of a two-span reinforced concrete (RC) bridge with \emph{shape memory alloy (SMA)-restrained rocking (SRR)} columns through machine learning (ML) techniques. SRR columns incorporate a combination of replaceable superelastic NiTi (SMA) links and mild steel energy-dissipating links to achieve self-centering and energy dissipation, respectively, while their rocking joints are protected against compressive concrete damage through steel jacketing. To produce seismic fragility functions, initially, \emph{multi-parameter} probabilistic seismic demand models (PSDMs) are generated for various engineering demand parameters through five different ML techniques (including neural network) and considering various sources of uncertainty, and the most accurate PSDMs are selected. The selected PSDMs are then interpreted using four different methods to investigate the effects of two key SRR column design parameters (self-centering coefficient and SMA link initial strain) and ambient temperature on the seismic performance of SRR columns. Subsequently, using neural networks, the PSDMs developed earlier, and appropriate capacity models, \emph{multi-parameter} fragility functions are developed for various bridge damage states. After examining the effects of the two SRR column design parameters on the seismic fragility of the bridge, its seismic fragility is compared with those of the same bridge with monolithic RC and posttensioned (PT) rocking columns. It is shown that, in general, increasing the initial strain of the SMA links and decreasing the self-centering coefficient as possible (i.e., without compromising the self-centering) reduce the overall bridge damage. In addition, even considering the ambient temperature's uncertainty, SRR columns are proven, at least, as effective as PT columns in mitigating the seismic damage of the bridges of monolithic RC columns.

\end{abstract}




\maketitle

\section{Introduction}\label{intro}
Bridges are among the key elements of transportation networks, making their continuous functionality crucial to the socioeconomic prosperity of modern communities. In order to effectively and efficiently manage and maintain the bridges in seismic regions, seismic risk assessments are necessary~\citep{chang2000probabilistic, kiremidjian2007seismic, padgett2007bridge, padgett2010regional,
kameshwar2014multi}. Seismic fragility functions, which determine the probabilities of exceeding various damage states (DSs) based on ground motion and structural characteristics~\citep{shinozuka2000statistical, shinozuka2000nonlinear, choi2004seismic, nielson2007analytical, padgett2008methodology}, are prerequisites to the seismic risk assessment of bridges. In addition to their paramount role in seismic risk assessment, fragility functions can provide great insights into the seismic performance of bridges (globally) and their components (locally), and thus, they can be used to probabilistically evaluate the seismic damage mitigation benefits of innovative bridge systems in comparison with their conventional counterparts~\citep{billah2012seismic, muntasir2015seismic, billah2018probabilistic, todorov2021seismic, salkhordeh2021seismic, fan2021machine, lu2022seismic}. Building upon the previous efforts on the seismic design of the so-called \emph{Shape Memory Alloy (SMA)-Restrained Rocking (SRR)} columns and the deterministic seismic performance assessment of the bridges equipped with those columns~\citep{akbarnezhad2022seismic,akbarnezhad2022bridge}, this paper aims to, first, evaluate the seismic vulnerability of the bridges of SRR piers by developing multi-parameter fragility functions using advanced machine learning (ML) techniques; and second, compare their seismic vulnerability with those of the bridges of posttensioned (PT) and monolithic reinforced concrete (RC) columns.







\section{Background}\label{background}
\subsection{SRR bridge columns}\label{SRR}
\begin{figure}
	\centering
		\includegraphics[scale=0.92]{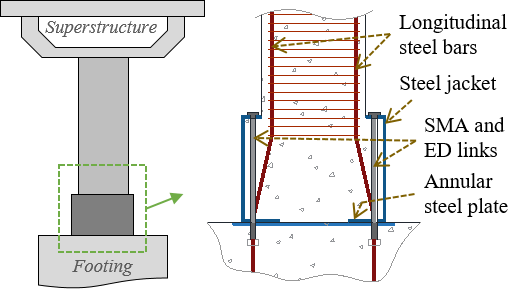}
	  \caption{Illustration of an SRR column with bottom rocking joint (design variation 1 per~\cite{akbarnezhad2022seismic})}\label{SRR_joint}
\end{figure}

SMAs are a class of “smart” materials capable of recovering their original shape after relatively large deformations (up to 6-8\% of strain), which is referred to as \emph{superelasticity} effect~\citep{hodgson1990shape,lagoudas2008shape}. The flag-shaped hysteretic behavior of SMAs has made them particularly attractive for the development of seismically-resilient structural components that would benefit from self-centering capabilities~\citep{desroches2004shape, ozbulut2011seismic}, including innovative bridge columns~\citep{wang2005study, saiidi2006exploratory, billah2012sma, moon2015seismic, tazarv2015reinforcing, tazarv2016low, varela2016bridge, varela2017resilient, tazarv2020analysis}. SRR columns, which were recently proposed within this context~\citep{akbarnezhad2022seismic}, are precast concrete columns with end steel-armored rocking joints controlled through two series of unbonded links: (\textit{i}) prestressed SMA links made of superelastic nickel-titanium (NiTi) alloy, and (\textit{ii}) energy-dissipating (ED) links made of mild steel (Fig.~\ref{SRR_joint}). The SMA links provide SRR columns with self-centering, while the ED links supplement the moderate energy dissipation provided by the SMA links. SRR columns are designed in a performance-based manner with the objective of sustaining minimal damage and residual displacement under the Maximum Considered Earthquake (MCE) hazard~\citep{akbarnezhad2022seismic}. Compared to the rocking columns posttensioned through internal high-strength steel tendons, SRR columns offer higher durability (due to the excellent corrosion resistance of NiTi) and easier repair (due to the easy replacement of SMA links).

In a recent study~\citep{akbarnezhad2022seismic}, three SRR column design variations were introduced and evaluated under cyclic lateral loading through 3D nonlinear continuum-based finite element models. The simulation results demonstrated that all three variations were successful in achieving the desired performance objectives, particularly, preventing concrete spalling and longitudinal rebar yielding over the unjacketed lengths of the columns and limiting their residual drift ratio to less than 0.3\% after being subjected to MCE displacement demands. More recently, \cite{akbarnezhad2022bridge} evaluated the seismic performances of two reinforced concrete (RC) bridges equipped with SRR columns through multiple nonlinear time history analyses and considering three discrete hazard levels. The results of this study showed that: (1) when properly designed, SRR columns are capable of achieving their targeted seismic performance objectives (i.e., damage and residual displacement prevention under MCE hazard); (2) vertical excitation and near-fault ground motions have inconsequential effects on the seismic damage of SRR columns; and (3) ambient temperature has very limited effect on the performance of SRR columns during earthquakes.

\subsection{ML-based seismic fragility assessment}\label{ML_based}
As mentioned earlier, seismic fragility functions are intended to compute the exceedance probabilities of various DSs given ground motion and structural characteristics~\citep{basoz1999statistical, mander1999seismic, shinozuka2000statistical}. In the literature, two approaches have been generally taken to develop analytical fragility functions, which are herein referred to as \emph{one-stage} and \emph{two-stage}. In both approaches, fragility functions are developed based on the exceedance probabilities that are obtained through the comparison of engineering demand parameters (EDPs), which represent various local/global structural responses, with respective capacities. However, the EDPs used in the two approaches are determined differently. In the one-stage approach, the EDPs are directly obtained from nonlinear time history analyses, necessitating a large data set~\citep{shinozuka2000statistical, shinozuka2000nonlinear, der2002seismic, porter2007creating, zhang2009evaluating, baker2011fitting, baker2015efficient, alam2012seismic, kiani2019application, salehi2020assessing, ghosh2021seismic, wang2022deep, rajkumari2022fragility}. In the two-stage approach, which is more commonly used than the one-stage approach, the EDPs are estimated via the so-called \emph{probabilistic seismic demand models} (PSDMs), which are initially generated based on the data obtained from the nonlinear time history analyses~\citep{choi2004seismic, nielson2007seismic, nielson2007analytical, padgett2008methodology, ramanathan2010analytical}. The two-stage approach could be helpful especially when the data set obtained from the nonlinear time history analyses is not sufficiently large. However, it could potentially introduce additional error (epistemic uncertainty) into the resulting fragility functions~\citep{der2009aleatory, zhang2009evaluating,  muntasir2015rev, bakalis2018seismic, zhang2020seismic, flenga2021fragility, conde2022seismic}.

Even though the common approach to develop PSDMs is using linear regression~\citep{xiang2019comparative, xiang2020probabilistic, wei2020evaluation, chen2021seismic}, which facilitates both model training and interpretation, the resulting linear regression models could not always represent the seismic response of structures with a high accuracy~\citep{ghosh2013surrogate, rokneddin2014seismic, mangalathu2018critical}. Moreover, linear regression models become even less effective/accurate if more than one independent variable are of interest in the PSDM development (i.e., not only a single IM)~\citep{dukes2013application, jeon2019parameterized}. Another limiting assumption within the traditional framework is that the probability distributions of different DSs have to be assumed (often, log-normal) resulting in fragility functions of low accuracy~\citep{noh2015development}. Given the foregoing and due to the significant predictive power offered by ML models, in recent years, ML techniques have been used for the development of fragility functions~\citep{ soleimani2022state, xie2020promise, soleimani2022probabilistic}. ML techniques allow developing more accurate PSDMs by avoiding the simplifying assumptions adopted by other approximate methods.~\citep{lagaros2007, jia2022structural}. While single-parameter PSDMs are primarily employed to develop fragility functions, ML-based multi-parameter PSDMs could be used for several other purposes, too, such as fast seismic performance assessment of bridges after earthquakes~\citep{mangalathu2019rapid, xu2020real, todorov2022post}, seismic design optimization of bridges~\citep{dukes2018development, fan2021machine}, and probabilistic investigation of effects of various geometric/material/design parameters on the seismic performance of bridges~\citep{moradi2018parameterized, mangalathu2019rapid, jeon2019parameterized}, especially innovative systems~\citep{shi2022seismic}.

Among the different ML techniques, several previous studies have used artificial neural networks to develop PSDMs for different types of structures~\citep{jia2022structural}. With the primary objective of reducing the computational cost of structural analysis, \cite{lagaros2007} proposed the rapid evaluation of seismic demands in steel building frames using a neural network. \cite{huang2020predicting} proposed an artificial neural network (ANN)-based capacity prediction model in order to take account of the uncertainty of the seismic performance of the reinforced concrete (RC) bridge columns. In order to eliminate the need for grouping the bridge classes based on their geometric characteristics (e.g., number of spans, number of columns per bent, and skew angle), \cite{mangalathu2018artificial} employed neural network (for PSDM development) and LASSO logistic regression (for fragility function development) to produce multi-parameter bridge-specific fragility functions. Other ML techniques have also been used for the development of fragility functions while studying the significance of parameter contributions to the output. In a study by~\cite{ghosh2013surrogate}, to develop bridge fragility functions, four different surrogate models (prediction models), namely, polynomial response surface, multivariate response adaptive regression spline, radial basis function networks, and support vector machines were explored and it was found that metamodels are capable of adequately and efficiently predicting bridge-specific seismic reliabilities with low computational cost. \cite{jeon2019parameterized} used multiple linear regression (for PSDM development) and logistic regression (for fragility function development) to develop bridge-specific fragility functions and showed that the stepwise removal process via a Bayesian parameter estimation method facilitates identifying the significant uncertain parameters. For the same purpose, \cite{mangalathu2019stripe} proposed using random forest, while their proposed method facilitated updating the fragility functions for new sets of input parameters. \cite{kiani2019application} also explored the application of one-stage classification-based ML algorithms in deriving seismic fragility curves for maximum inter-story drift ratio of an 8-story special steel moment resisting frame building and examined the significance of different IMs on the output based on LASSO and logistic methods. To predict seismic response of bridges against lateral spreading, \cite{wang2021seismic} examined five ML techniques, namely, multiple linear regression, LASSO regression, neural network, random forests, and gradient boosting and compared their performance using the coefficient of determination, $R^2$, and performance plot. Moreover, they tried to identify the most significant IM and explored the relative importance of the input variables obtained from the white-box models (all except neural network, which is deemed a black-box model). \cite{soleimani2022bridge} developed a framework for probabilistic seismic resilience assessment of common highway bridges using ensemble learning-based predictive, namely, bagging and boosting techniques. The final output could predict resilience index as a function of seismic events and bridge characteristics.

In addition to the high predictive power of the ML-based prediction models, interpretation of these models would provide insight into the mechanics of the problem~\citep{mangalathu2020failure, sun2021machine}. Interpretation of ML models refers to understanding the general relationships governing the data that are learned by the models~\citep{murdoch2019definitions}. Interpretation/inspection of ML models could help to further validate the trained models and to understand the influence of the input features on the model predictions. Some ML models are easy to interpret based on the trained model itself, such as linear regression or decision tree. However, some other ML models cannot be interpreted as easily because the trained models are more or less "black boxes". To overcome this limitation, there exist several interpretation methods that can be used to study the ML models that are not immediately interpretable, e.g., neural network. Some of these methods are model-specific, i.e., they can be used to interpret/inspect only specific ML models. However, there exist other methods that can be applied to any ML models, which are referred to as \emph{model-agnostic}~\citep{lipovetsky2001analysis, chakraborty2017interpretability, lundberg2017unified, guidotti2018survey, molnar2020interpretable, molnar2020interpretable2}.~\cite{soleimani2021analytical} assessed the importance of modeling parameters on seismic demand estimation for concrete box-girder bridges with tall piers using Random Forest ensemble learning method. Recently, \cite{mangalathu2022machine} used such interpretation methods to explore the key influential variables and thresholds governing the ML-based seismic demand prediction models for reinforced concrete bridges.

\section{Scope and objectives}\label{scope}
Although the overall seismic damage mitigation capability of SRR columns has been demonstrated through cyclic analyses~\citep{akbarnezhad2022seismic} and time history analyses~\citep{akbarnezhad2022bridge} representing certain hazard levels, the seismic vulnerability of bridges with SRR columns considering various uncertainties remains unexplored. Moreover, to show that SRR columns can seismically perform, at least, as well as PT columns, yet outperform monolithic RC columns, comparative fragility assessments are necessary. To address these points, this study aims to:
\begin{itemize}
\item Develop multi-parameter PSDMs for a highway bridge with SRR columns using a number of different ML techniques and choose the optimal PSDMs among them;
\item Investigate the effects of various input parameters (including two key SRR column design parameters and ambient temperature) on the seismic performance of the bridges with SRR columns by interpreting the chosen PSDMs via a few different methods;
\item Develop multi-parameter fragility functions for the bridge with SRR columns using ML techniques;
\item Examine the effects of SRR column design parameters on the fragility curves of the bridge with SRR columns;
\item Compare the seismic fragility of the bridge with SRR columns with those of the same bridge but with monolithic RC columns and PT columns.
\end{itemize}

The rest of this paper is organized as follows. The ML techniques considered in this study for PSDM development are briefly reviewed in Section~\ref{overview}. Section~\ref{CaseBridge} describes the highway bridge studied in this paper. Section~\ref{design} is concerned with the design of the SRR columns incorporated in the bridge and their equivalent PT columns. The numerical modeling of the selected bridge with different column types is described in Section~\ref{Numerical}. The PSDMs for the bridge with SRR columns are developed and interpreted in Section~\ref{PSDM}. The fragility functions of the bridges with monolithic RC columns, PT columns, and SRR columns are developed and compared in Section~\ref{frag_anal}. The effects of design parameters on the fragility functions of the bridge of SRR columns are examined in the same section. Finally, Section~\ref{conclusions} provides the summary and conclusions.

\section{Overview of ML techniques}\label{overview}
Five different ML techniques were considered to develop the PSDMs of the bridge with SRR columns. These techniques, which include (1) ridge regression, (2) kernel support vector regression, (3) adaptive boosting, (4) random forest, and (5) neural network, are briefly reviewed in this section. The fundamentals and the formulations of the selected ML techniques are not extensively discussed here for brevity and they can be found elsewhere~\citep{bishop2006pattern, hastie2009elements, murphy2012machine}.

\subsection{Ridge regression}
Ridge regression is basically a regularized version of multiple linear regression introduced by~\cite{hoerl1970ridge}. To reduce the model variance and complexity (i.e., "overfitting"), especially in the presence of multicollinearity (when some of predictors are correlated), in the ridge regression, a penalty term is added to the ordinary least squares loss function, i.e., sum of squared residuals (SSR). This penalty term is the sum of the squares of the regression coefficients multiplied by a regularization constant. Accordingly, by minimizing the penalized sum of squared residuals (PSSR) as its loss function, the ridge regression tends to make the coefficients corresponding to less significant predictors smaller, thereby improving the resulting model's stability and eliminating multicollinearity. Though an inherently nonlinear response cannot be accurately represented by a ridge regression model using the original predictors, expanding the original predictor space to include higher-order features (i.e., increasing the model dimensions) could improve the model's performance. However, the required feature expansion results in a large number of features, making the training process practically inefficient.

\subsection{Kernel support vector regression}
Similarly to multiple linear regression, support vector regression methods predict the target variable through a hyperplane. However, to achieve lower variance and better stability, the basic loss function in the support vector regression is defined as one half of the sum of the squares of the hyperplane coefficients, with the prediction errors being constrained within a certain margin~\citep{vapnik1999nature}. Since the support vector regression cannot accurately represent a nonlinear target using the original predictors (without transformations), kernel support vector regression is employed. In this technique, the loss function remains the same, but the predictors are transformed via kernel functions before entering the hyperplane equation~\citep{boser1992training}. The kernel functions generally quantify the similarities (e.g., distances) between the observations. In this study, the Gaussian kernel function, which is also referred to as Radial Basis Function, was used. It is noted that the kernel support vector regression is computationally more efficient than the ridge regression for learning nonlinear responses because it does not require the original space to be actually transformed into higher dimensions. 

\subsection{Random forest}
Ensemble ML technique have also been developed to improve the stability/robustness of ML-based prediction models~\citep{sagi2018ensemble}. These techniques combine the predictions of a large number of base estimators ("weak learners") to make a final prediction. In the context of regression, the final prediction is the (weighted) average of the predictions of the base estimators. Random forest is an ensemble technique whose base estimators are low-depth decision trees, intending to reduce the overall model's variance compared to a single deep decision tree~\citep{breiman2001random}. To this end, each decision tree is trained using only a subset of the entire training data set, which is selected randomly with replacement (known as "bootstrap" sampling or "bootstrapping"), while the features considered for splitting at each decision node are also picked from a random subset of all the existing features. It is worth noting that random forest is a \emph{parallel} ensemble technique, i.e., each decision tree is trained independently of the other decision trees.

\subsection{Adaptive boosting}
Adaptive boosting, which is more often referred to as "AdaBoost", is also an ensemble technique with similar objectives to the random forest's~\citep{freund1997decision}. However, AdaBoost is of a \emph{sequential} type (not parallel), i.e., its base estimators are trained one after another and considering the error in the previous training. That is, each estimator is trained using the entire data set, but the weights of the observations (used in the loss function) are adjusted at each training step on the basis of the prediction errors of the previous trained estimator. As a result of this iterative process, the base estimators gradually become stronger in making more difficult predictions. In this study, the base estimators used in the AdaBoost method were low-depth decision trees. 

\subsection{Neural network}
Neural networks loosely emulate human brains and are capable of predicting the most complex nonlinear responses~\citep{haykin2009neural}. Each neural network consists of a series of neurons organized in several layers, with the first and the last layers including the input and output variables, respectively, while each neuron has a weight associated with it. The basic form of neural network (only including input and output layers) represents a multiple linear regression model. Adding intermediate layers (including "hidden" neurons) can make the model nonlinear. Before the outputs of each layer are given to the next layer as its inputs, they are transformed using an "activation function" to increase the model's predictive power and robustness. In this study, the rectified linear unit (ReLU)~\citep{goodfellow2016deep} and linear functions are used as the activation functions for the hidden layers and the output layer, respectively. 


\section{Bridge description}\label{CaseBridge}
\begin{figure}
	\centering
	\includegraphics[scale=0.72]{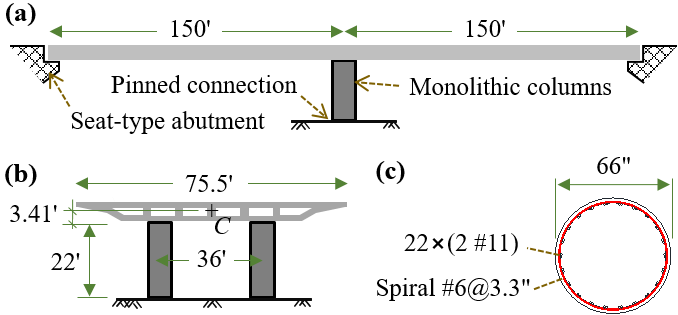}
	\caption{La Veta Avenue overcrossing bridge: (a) bridge elevation; (b) bent elevation; (c) column cross section}
	\label{laveta}
\end{figure}

The La Veta Avenue overcrossing bridge located in Orange, California, was selected for the seismic fragility assessments. The key features of the bridge are illustrated in Fig.~\ref{laveta}. The bridge is made of RC and it has two spans with seat-type abutments and a single two-column bent. The superstructure consists of a continuous six-cell RC box-girder, which spans the total length of 299.8~ft and width of 75.5~ft. For simplicity, the bridge's two spans were assumed to be identical (Fig.~\ref{laveta}(a)). The superstructure is supported by seven elastomeric bearing pads at each abutment. The bent consists of two identical columns, each being 22~ft tall, and a cap-beam integral with the bridge deck (Fig.~\ref{laveta}(b)). The cross section of the original columns is 66~in. in diameter (Fig.~\ref{laveta}(c)) and each column has an axial load ratio of about 10\%. The original monolithic RC columns are of pinned connections at their bottom ends. The concrete's nominal compressive strength, $f'_c$, is 5~ksi and the reinforcing steel is of Grade 60~\citep{astm2009standard}.  Further details of the bridge can be found in~\cite{kaviani2014}.

\section{Design of rocking columns}\label{design}
This section summarizes the designs of the SRR and PT columns that would replace the original monolithic RC columns of the selected bridge. For a fair comparison in the subsequent seismic fragility assessments (Section~\ref{col_frag_comp}), the rocking columns' boundary conditions were assumed to be the same as the monolithic columns' (fixed at their top ends and pinned at their bottom ends) and one of the design objectives was to achieve nearly equal maximum base shear resistances in all the piers, i.e., with monolithic columns, PT columns, and SRR columns.

\subsection{SRR columns}
The SRR columns were selected to be of design variation~1 per~\cite{akbarnezhad2022seismic} (see Fig.~\ref{SRR_joint}). A performance-based approach was taken to design the SRR columns, while the SRR pier's maximum lateral load resistance was sought to remain equal to that of the monolithic pier (see Fig.~\ref{push_comp}). The broad performance objective was to prevent the SRR columns from repair-necessary damage and residual deformations under an MCE (2475-year) event. More specifically, the following DSs were aimed to be prevented under the MCE displacement demand: (1) concrete spalling/crushing, (2) longitudinal rebar yielding, (3) strength deterioration or fracture of ED links, and (4) plastic deformation of SMA links. The SRR columns were designed through an iterative process~\citep{akbarnezhad2022seismic} until the above performance objectives were satisfied. Here, the MCE displacement demand was determined using the capacity spectrum method~\citep{chopra1999capacity, fajfar1999capacity, madhusudhanan2018capacity}, while the requisite nonlinear static (pushover) analyses were conducted via the numerical model described subsequently - the final MCE drift ratio demand was 3.6\%.

\begin{figure}
    \centering
    \includegraphics[scale=0.50]{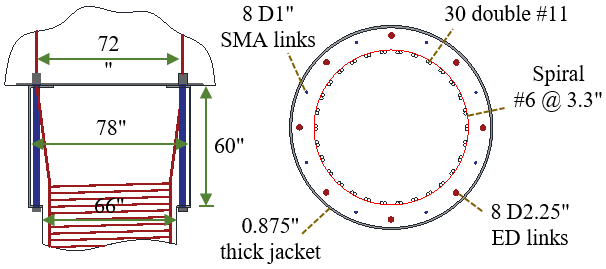}
    \caption{Design details of SRR columns}
    \label{ColDim}
\end{figure}

The dimensions and the reinforcement details of the designed SRR columns are shown in Fig.~\ref{ColDim}. The diameter of the columns over their unjacketed lengths was selected to be identical to the diameter of the original monolithic RC columns (66 in.). To accommodate the SMA and ED links, the diameter of the columns' jacketed cross section was selected to be 12 in. larger. The longitudinal reinforcement of each SRR column consists of 30 double \#11 bars, resulting in a longitudinal reinforcement ratio of 2.7\%. The transverse reinforcement over the unjacketed length was kept similar to the original monolithic RC columns'. The steel jacket was assumed to be made of A36 steel~\citep{astm2014standard}, 7/8 in. thick, and 60 in. tall. Totals of eight SMA links and eight ED links (also made of A36 steel) are used at the rocking joint of each column. The SMA and ED links are 60 in. and 45 in. long, respectively, and their cross-sectional areas are 0.79 in.$^2$ and 3.97 in.$^2$, respectively.

\subsection{PT columns}
\begin{figure}
	\centering
		\includegraphics[scale=0.50]{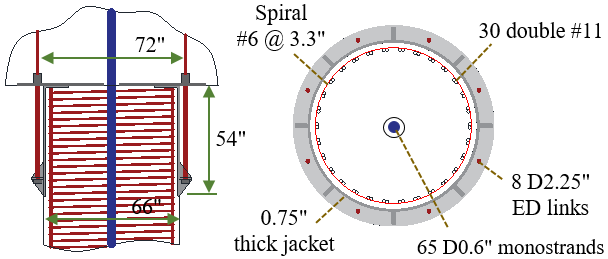}
	  \caption{Design details of PT columns}\label{PT_det}
\end{figure}

\begin{figure}
	\centering
		\includegraphics[scale=0.90]{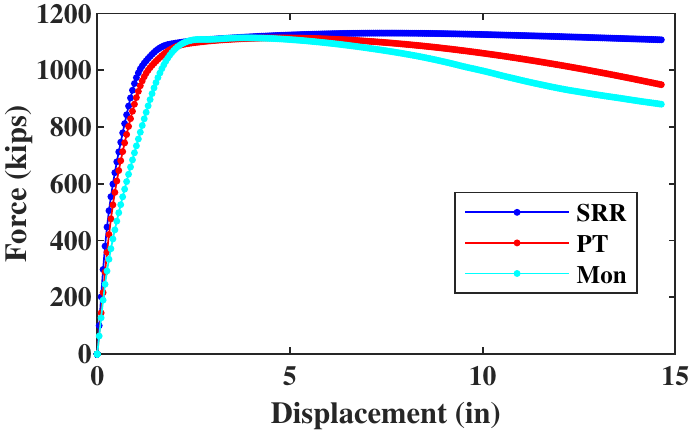}
	  \caption{Comparison of pushover curves of three different bridge piers}\label{push_comp}
\end{figure}

The PT columns were chosen to be of central unbonded tendons and their rocking joints were steel-armored (Fig.~\ref{PT_det}). The same amount of ED links as in the SRR columns was provided in each of the PT columns. The PT columns were selected to be of the same diameter as the monolithic columns', i.e., 66 in. Once these parameters were fixed, the total area of the unbonded PT tendon and their initial posttensioning force were iteratively selected such that the PT pier's lateral load resistance got reasonably close to those of the monolithic and SRR piers (Fig.~\ref{push_comp}). The selected PT tendon (Grade 270 per~\cite{astm2016standard}) consisted of 65 0.6-in. dia. monostrands and its total initial PT force was 1,205 kips (equal to 35\% of its total yield strength). The height of the steel jacket, which was assumed to be made of A36 steel~\citep{astm2014standard}, was selected as 54 in. to avoid concrete spalling under the pier's MCE displacement demand.

\section{Numerical model}\label{Numerical}
\begin{figure*}
	\centering
		\includegraphics[scale=0.45]{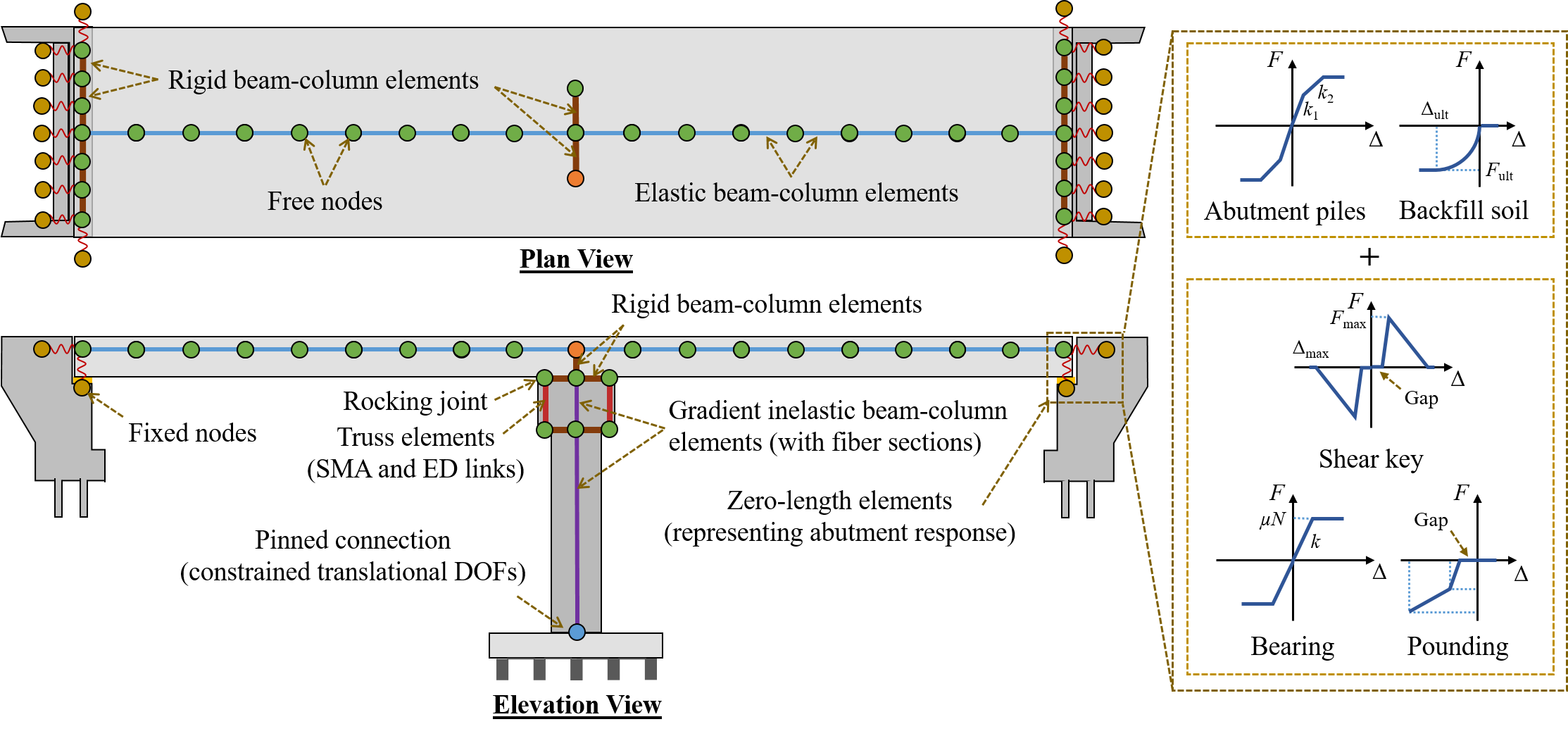}
	  \caption{Finite element modeling of bridge with SRR columns}\label{BridgeNum}
\end{figure*}

The 3D model of the bridge was developed through the finite element analysis program OpenSees~\citep{mckenna2011opensees}. The schematic of the numerical model is presented in Fig.~\ref{BridgeNum}. Major parts of the model are described in the separate sections below.

\subsection{Superstructure and abutments}
Elastic beam-column elements were used to model the superstructure's longitudinal spine as it was expected to remain elastic under earthquake excitation. At the two ends of the bridge, five series of zero-length elements were used to represent external shear keys (in the transverse direction), pounding (in the longitudinal direction), backfill soil (in the longitudinal direction), bearings (in the both transverse and longitudinal directions), and abutment piles (in the both transverse and longitudinal directions). Nearly rigid elastic elements were used to connect the above zero-length elements to the superstructure's spine elements. The zero-length elements representing the shear keys and bearings were connected in series to the zero-length elements representing the abutment piles, while the  zero-length elements representing the pounding were connected in series to the zero-length elements representing the abutment's backfill soil. More details on the responses of these zero-length elements can be found elsewhere~\citep{mangalathu2016ancova, mangalathu2017performance, muthukumar2006hertz}.

\subsection{Columns}
\paragraph{Monolithic RC columns.}
Each monolithic column was modeled via a single gradient inelastic (GI) force-based beam-column element~\citep{sideris2016gradient, salehi2018finite, salehi2020assessing}. The GI element's characteristic length, $l_c$, was taken as the column's plastic hinge length calculated per~\cite{goodnight2016modified}. A total of 13 integration points were considered along each element to ensure objective responses upon softening~\citep{sideris2016gradient, salehi2017refined}. Each integration point was represented by a fiber section consisting of uniaxial material models simulating unconfined concrete, confined concrete, and longitudinal steel. The adopted uniaxial material models for concrete and reinforcing steel were the Kent-Scott-Park model~\citep{scott1982stress} and the Giuffre-Menegotto-Pinto model~\citep{giuffre1970behavior}, respectively. The behavior of the confined concrete was determined based on the Mander's model~\citep{mander1988theoretical}.

\begin{figure}
	\centering
		\includegraphics[scale=0.75]{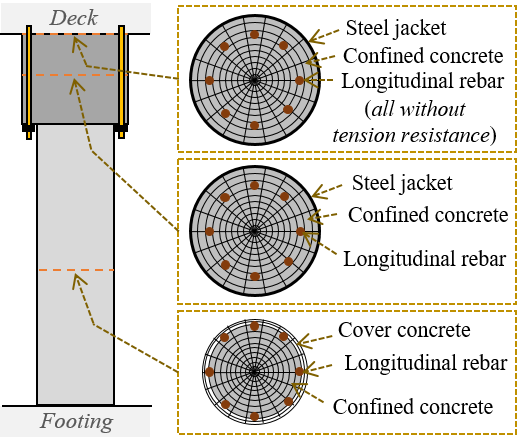}
	  \caption{Fiber sections of beam-column elements representing SRR columns}\label{ColNum}
\end{figure}

\paragraph{PT columns.}
Each PT column was modeled via two GI elements representing the jacketed and unjacketed lengths of the precast concrete segment, a multi-node co-rotational continuous truss element representing the central tendon, and multiple zero-length gap elements simulating the duct-tendon interactions at the joints ~\citep{salehi2020effect}. Three different fiber sections were used in the GI elements to represent the different regions of the column, i.e., its rocking joint, its jacketed length, and its non-jacketed length (e.g., see Fig.~\ref{ColNum}). The jacketed length's fiber sections did not have unconfined concrete regions and were enclosed by steel fibers representing the steel jacket. The rocking joint's fiber section was similar to the jacketed length's, except its material models did not resist tension. The GI element's characteristic length, $l_c$, was selected as the cross section radius. The ED links were simulated via co-rotational truss elements, which were connected to the column (at the node between the jacketed and unjacketed lengths) through nearly rigid elastic beam-column elements. The multi-node truss element consisted of four nodes, one at the anchorage point inside the footing, one at the footing-column joint location (connected through gap elements to fixed nodes at the same location), one at the column-superstructure joint location (connected through gap elements to the nodes that were connected to the superstructure's spine elements), and the last one at the anchorage point inside the superstructure (connected to the superstructure's spine elements). The posttensioning steel was represented by a tension-only material model following the backbone curve suggested by Mattock~\citep{mattock1979flexural} and capable of modeling the post-plasticity slackness of the strands~\citep{salehi2017numerical}. The Giuffre-Menegotto-Pinto model was used to simulate both the ED links and the steel jacket, while it was combined with a low-cycle fatigue model~\citep{uriz2005towards} in case of ED links to allow evaluating their cyclic damage after earthquakes.

\begin{figure*}
	\centering
	\includegraphics[scale=0.25]{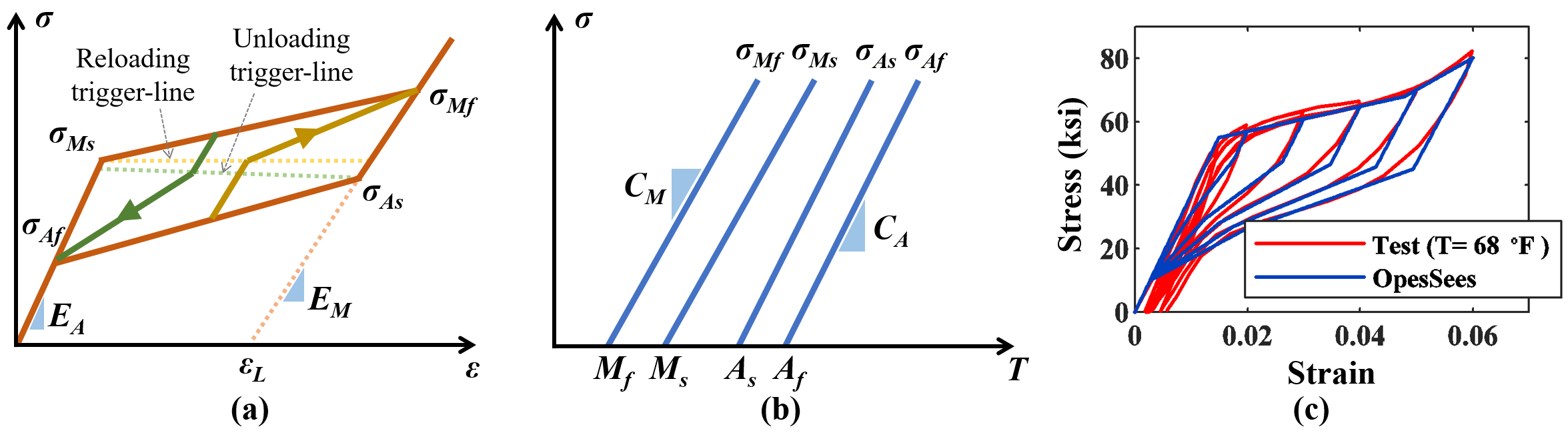}
	\caption{Superelastic material model: (a) illustration of trigger-line model; (b) temperature dependence; (c) calibration per \cite{desroches2004cyclic}
	}\label{SMA_validation}
\end{figure*}

\paragraph{SRR columns.}
The element configuration used to model SRR columns can be seen in Fig.~\ref{BridgeNum}. The height of each SRR column was discretized into two GI beam-column elements representing the jacketed and unjacketed lengths of the column and three different fiber sections were used to represent the different integration points (Fig.~\ref{ColNum}). Similarly to the PT columns, the fiber section representing the integration point at the rocking joint location consisted of compression-only material models. The intermediate node (between the two GI elements) was necessary to connect the co-rotational truss elements representing the SMA and ED links to the GI elements. The characteristic lengths of the GI elements, $l_c$, were selected as the radii of their respective cross sections. The concrete and steel material models used for modeling the SRR columns were similar to those used to model the PT columns. The superelastic NiTi was represented by the superelastic uniaxial material model, which is described subsequently. This model was calibrated based on the cyclic response of a 1-in. diameter superelastic NiTi rod tested by \cite{desroches2004cyclic} (Fig.~\ref{SMA_validation}(c)).

\paragraph{Superelastic material model.}
None of the uniaxial material models readily available in the material library of OpenSees could realistically capture the temperature dependence and the sub-cycles of the superelastic NiTi's hysteresis. As a result, for the simulation of SMA links in the SRR columns, a new material model, termed \emph{superelastic}, was developed and implemented in OpenSees. To capture the hysteresis sub-cycles, the superelastic model follows the basic loading/unloading rules of the trigger-line model proposed by~\cite{muller1991pseudo} - see Fig.~\ref{SMA_validation}(a). Moreover, as shown in Fig.~\ref{SMA_validation}(b), the transformation stresses of the material are considered to be linearly dependent on the ambient temperature~\citep{brinson1993one}. It is worth noting that, assuming that the superelastic NiTi bars used to make the SMA links are appropriately heat-treated, they are assumed to experience negligible residual/plastic deformations~\citep{salehi2022experimental}. Therefore, the developed superelastic material model considers full superelasticity. 

\section{Probabilistic seismic demand models}\label{PSDM}
This section aims to: (1) explain the process of developing several PSDMs for the bridge with SRR columns using five different ML techniques (see Section \ref{overview}); (2) select the most efficient ML technique for developing the PSDMs; and (3) use various interpretation methods to explore the final developed PSDMs.

\subsection{Design of experiment}
A suite of ground motions selected by~\cite{baker2011new} was used in this study to account for the ground motion uncertainties. The original suite included 160 ground motion records consisting of 120 far-field and 40 near-fault ground motions. However, here, the suite was expanded to 320 records by scaling all the original 160 ground motions by a factor of 2 in order to include sufficiently strong ground motions~\citep{ramanathan2012next}. In addition, the uncertainties in the nonlinear behavior of the SRR columns due to the variations of material properties and ambient temperature, $T$, were included in the time history analyses. Table~\ref{table_input} summarizes the parameters of the distributions considered for different material properties. The coefficients of variation for concrete and steel properties were selected in accordance with~\cite{mangalathu2017performance}, whereas the coefficients of variation for NiTi properties were selected based on~\cite{gur2021design}. In addition, Table~\ref{table_input} includes the considered range of ambient temperature for the bridge location according to~\cite{aashto2012aashto}. Two design parameters of SRR columns, namely, self-centering coefficient, $\lambda$, and initial strain of SMA links, $\epsilon_{SMA,0}$~\citep{akbarnezhad2022seismic}, were also assumed to be randomly generated via uniform distributions with the parameters listed in Table~\ref{table_input}. According to \cite{akbarnezhad2022seismic}, $\lambda$ values of 0.0 and 1.0 correspond to the SRR column designs with the maximum possible ED links to maintain self-centering and without any ED links, respectively. Including the design parameters and the ambient temperature among the uncertain variables would allow exploring the effects of these parameters on the performance of the bridge with SRR columns subsequently. Latin Hypercube sampling was utilized to generate 320 sample pairs of ground motion and bridge model with randomly selected values for the variables of Table~\ref{table_input}. Subsequently, a nonlinear time history analysis was conducted for each sample to obtain the EDPs of interest (see Section~\ref{EDPs}).  

\begin {table*}[h!]
\centering
\caption{Considered distributions for different uncertain parameters; min. and max. values for normal and log-normal distributions correspond to the lower and upper truncated limits, respectively.}
\label{table_input}
\begin{tabular}{ p{6cm}p{1.cm}p{2cm}p{1.2cm}p{1.5cm}p{1.2cm}p{1.2cm}}
\hline

 \multirow{2}{*}{Parameter} & \multirow{2}{*}{Unit} & \multicolumn{5}{c}{Statistical Distribution}\\
  \cline{3-7}
  &  & Distribution & Mean & COV (\%) & Min. & Max.\\ 
\hline
Concrete compressive strength, $f'_c$ & ksi & Normal & 4.55 & 12.0 & 3.46 & 5.64\\
&&&&&&\\
Rebar yield strength, $f_{y,s}$ & ksi & log-normal & 69.0 & 8.0 & 58.8 & 81.0 \\
Steel jacket yield strength, $f_{y,j}$ & ksi & log-normal & 44.0 & 8.0 & 37.5 & 51.6\\
ED link yield strength, $f_{y,ED}$ & ksi & log-normal & 48.0  & 8.0 & 40.9 & 56.3\\
&&&&&&\\
NiTi martensite start stress, $\sigma_{Ms}$ & ksi & log-normal & 55.0 & 10.0 & 45.0 & 67.2\\
NiTi martensite finish stress, $\sigma_{Mf}$ & ksi & log-normal & 82.0 & 10.0 & 67.1 & 100.2\\
NiTi austenite start stress, $\sigma_{As}$ & ksi & log-normal & 55.0 & 10.0 & 45.0 & 67.2\\
NiTi austenite finish stress, $\sigma_{Af}$ & ksi & log-normal & 20.0 & 10.0 & 16.4 & 24.4\\
NiTi transformation strain, $\epsilon_L$ & in./in. & log-normal & 0.035 & 5.0 & 0.03 & 0.04\\
NiTi Young's modulus of austenite, $E_A$ & ksi & log-normal & 3700.0 & 5.0 & 3347.9 & 4089.1\\
NiTi Young's modulus of martensite, $E_M$ & ksi & log-normal & 3300.0 & 5.0 & 2986.0 & 3647.1\\
&&&&&&\\
Ambient temperature, $T$ & $^\circ$C & Uniform & 25.0 & 46.2 & 5.0 & 45.0\\
&&&&&&\\
Initial SMA link strain, $\epsilon_{SMA,0}$ & \% & Uniform & 0.75 & 57.7 & 0.0 & 1.5\\
Self-centering coefficient, $\lambda$ & - & Uniform & 0.5 & 57.7 & 0.0 & 1.0 \\
\hline
\end{tabular}
\end{table*}

\subsection{Input features}
In general, ML techniques allow developing PSDMs that depend on multiple input variables/features, e.g., ground motion IMs and design parameters. Within the framework of seismic risk assessment, in order to obtain the exceedance probability of a decision variable (such as repair cost), fragility functions need to be integrated with respect to the ground motion IMs predicted by seismic hazard curves~\citep{Cornell2000progress, moehle2004framework}. Despite the fact that many researches have explored the possibility of using a vector-valued IM as a substitute for a single IM in risk assessment~\citep{tothong2007probabilistic, baker2005vector, gehl2013vector,kohrangi2016vector}, the seismic hazard curves dependent on a vector-valued IM are not easily available~\citep{bazzurro2002vector, abrahamson2006seismic, silva2019current}. As a result, fragility functions and PSDMs are usually conditioned on a single IM rather than multiple IMs. It is worth noting that in cases where only the PSDMs are of interest (e.g., for seismic response prediction), however, those PSDMs that are trained based on multiple IMs could potentially be more useful~\citep{xu2020real}. 

Since the selected IM may significantly affect the degree of uncertainty of the PSDMs, the IM selection is not a trivial task and has been the subject of several past studies~\citep{giovenale2004comparing, baker2008vector, tothong2007probabilistic, eads2015average, hariri2016probabilistic, zelaschi2019critical, du2019posteriori,guo2020optimal, du2021entropy}. The process of selecting an appropriate IM for the PSDMs of the bridge with SRR columns is explained in Section~\ref{secCurse}. However, 11 ground motion characteristics were considered herein as potential IMs - see Table~\ref{table:IM}. All the considered IMs are independent of the structural properties to ensure they are conveniently hazard-computable. Overall, possible input features for training the PSDMs included all the variables listed in Table~\ref{table_input} plus a single IM. However, not all of these features were used in the final trained models to ensure robustness and efficiency, as discussed in Section~\ref{secCurse}.

\begin{table}[]
\caption{Potential intensity measures for PSDM development}
\label{table:IM}
\begin{tabular}{ll}
\hline
IM          & \multicolumn{1}{c}{Definition}  \\ \hline
PGA         & Peak ground acceleration             \\
PGV         & Peak ground velocity                \\
PGD         & Peak ground displacement             \\
CAA         & Cumulative absolute acceleration             \\
CAV         & Cumulative absolute velocity             \\
CAD         & Cumulative absolute displacement              \\
CSA         & Cumulative squared acceleration (Arias intensity)      \\
CSV         & Cumulative squared velocity   \\
CSD         & Cumulative squared displacement \\
$Sa_1$      & Spectral acceleration at period of 1 sec. \\
$Sa_{gm}$   & Geom. mean of $Sa$ for period range of 0.5-2 sec.   \\ \hline
\end{tabular}
\end{table}


\subsection{Engineering demand parameters}\label{EDPs}
The EDPs of interest in this study represent both global and local responses. The selected global EDPs are the peak displacement demands at the deck ends in both transverse and longitudinal directions (to detect abutment unseating), the bent's peak drift ratio, and the bent's residual drift ratio. The selected local EDPs are the maximum tensile strains of the longitudinal rebar, the ED links, and the SMA links; the maximum compressive strains of the cover and core concrete; the low-cycle fatigue damage factors of the ED links; and the peak deformation demands of the shear keys and the bearing pads.

\subsection{Data preprocessing}
Before developing the PSDMs, both the input data and the predictions (i.e., the EDPs obtained via the time history analyses of the bridge with SRR columns) needed to be preprocessed. This step usually includes removing the potential outliers and variable transformations. As outliers, the data resulted from the incomplete analyses (i.e., without solution convergence) were removed - this included only a handful of cases. After the outlier treatment, the new data set was randomly split into training and test data sets. The training data set (here, comprising 70\% of the entire data) was used to train the PSDMs, while the test data set was used to measure the performance of the trained PSDMs. The variable transformations were carried out after splitting the data set to avoid "data leakage" from the test set to the training set. The variable transformations included the log-transformation of those EDPs that had skewed distributions and the normalization of the input features due to their different ranges. The former was done to improve the performance of the trained models and the latter was done to overcome the slow learning of certain ML models (e.g., using numerical optimization algorithms such as gradient descent~\citep{ruder2016overview}) and to avoid the dominance of the errors of large-valued feature in the loss function calculations. In this study, the input features were normalized via Z-score transformation.


\subsection{Hyperparameter tuning}
After the data preprocessing, any of the ML techniques described in Section~\ref{overview} could be used to train the PSDMs. Each ML model is trained through optimizing the associated objective function (e.g. cost function in neural network, information gain in decision tree). In addition, there are some other parameters that need to be specified prior to the training of each model, which are referred to as "hyperparameters". Examples of hyperparameters in the ML techniques considered herein include the regularization constant in the ridge regression, and the number of layers and the number of neurons within each layer in the neural network. Hyperparameter tuning is necessary to improve the performance of a trained model and avoid its overfitting. The hyperparamter tuning is usually achieved through grid/randomized search and cross-validation techniques~\citep{hastie2009elements}. The search methods simply seek the best combination of all hyperparameters (within prespecified ranges) that would maximize a selected cross-validated score (e.g., root mean square error and $R^2$). The main purpose of the cross-validation technique is to avoid overfitting, which would occur if the hyperparameters were merely selected based on the model's performance on the training data set. In this study, a five-fold cross-validation was used and $R^2$ was chosen as the performance metric that needed to be maximized through the hyperparameter tuning. 

\subsection{Dimensionality reduction}\label{secCurse}
The term "curse of dimensionality" refers to the model training difficulties arising from having many input features relative to the number of observations (data sparsity). This is also known as the Hughes phenomenon~\citep{hughes1968mean}, which states that the overall power of a prediction model initially increases with the number of features/dimensions, but it stops increasing at some point and even deteriorates as the number of features increases further. This occurs because the data sparsity precludes the ML model from generalizing to the regions of feature space that do not have sufficient data for model training. 

\begin{figure*}
	\centering
		\includegraphics[scale=1.0]{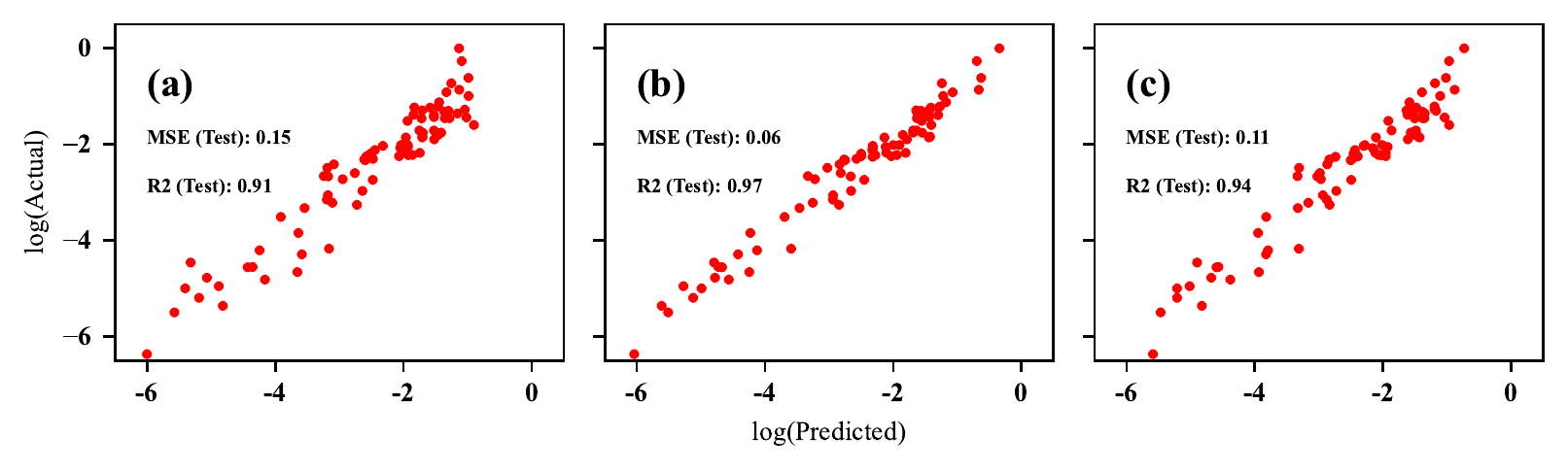}
	  \caption{Performance comparison of longitudinal drift prediction model with different features: (a) single IM; (b) single IM + design parameters + ambient temperature; (c) multiple IMs + material properties + design parameters + ambient temperature.}\label{CompPerf}
\end{figure*}

In order to explore how different number of dimensions could affect the model performance, Fig.~\ref{CompPerf} compares the performance plots showing the actual peak bent drift values versus their corresponding neural network predictions for three cases: (1) the only feature considered is $Sa_1$ (i.e., only one IM); (2) the considered features include the two SRR column design parameters ($\epsilon_{SMA,0}$ and $\lambda$), the ambient temperature ($T$), and $Sa_1$ (i.e., one IM plus only three extra dimensions); (3) the considered features include all the IMs in Table~\ref{table:IM}, the SRR column design parameters, the material properties, and the ambient temperature (i.e., all the possible uncertain parameters). It is observed that adding new features has increased the model's performance in case~2 ($R^2=0.97$) with respect to case~1 ($R^2=0.91$). However, in case~3, adding multiple extra features has decreased the model's performance compared to case~2 ($R^2=0.94$ vs. $R^2=0.97$). This is attributed in part to the lower variances of the material properties compared to the two design parameters ($\epsilon_{SMA,0}$ and $\lambda$) and the ambient temperature ($T$), thereby giving the former features less predictive power. The potential correlation between the multiple IMs considered in case~3 could also have contributed to the model's lower performance in that case in comparison with case~2. Given the foregoing, the PSDMs developed for the bridge with SRR columns were decided to be similar to case 2 (i.e., their input features only included the design parameters, ambient temperature, and one IM ). The single IM used in the PSDMs was selected as explained below.

\paragraph{Selection of optimal IM.}
The optimal IM should have the following characteristics~\citep{padgett2008selection}: (\textit{i}) \emph{hazard computablity}, i.e., the hazard curve of that IM (obtained through probabilistic seismic hazard analysis) should be conveniently available; (\textit{ii}) \emph{proficiency}, i.e., the IM shall have a high predictive power; and (\textit{iii}) \emph{sufficiency}, i.e., the errors/residuals of the prediction model are independent of other ground motion parameters, such as magnitude and epicentral distance. For some of the selected IMs (see Table~\ref{table:IM}), namely, PGA, PGV, PGD, and $Sa_1$, the hazard curves are readily available from the U.S. Geological Survey~\citep{USGS}. In addition, according to several studies~\citep{jalayer2012analyzing, ebrahimian2015preliminary, du2019posteriori}, sufficiency is closely related to proficiency, so it is not a matter of concern provided that the IM proficiency is checked. 

Here, the levels of proficiency (higher importance and lower dispersion) of the possible IMs were examined by comparing the permutation feature importance~\citep{breiman2001random} values of the different IMs when all were included in PSDM development. The permutation importance can be computed for any feature regardless of the ML technique. The permutation importance of a feature equals the decrease in the prediction model's performance metric of interest (e.g., $R^2$) upon removing that feature from the model's input variables. The permutation importance values for all of the potential IMs used to train different ML models to predict the bent's peak drift ratio are listed in Table~\ref{table:Perm_Imp}. As seen, the $Sa_1$ has the largest importance for all the techniques. A similar trend was observed for other responses (except residual drift and damage factor for low-cycle fatigue of the ED links), although not shown here. Based on the permutation feature importance for all the EDPs, the $Sa_1$ outperforms the other IMs, so it was selected as the most appropriate IM for this study.

\begin{table}[]
\caption{Permutation importance values of different IMs used to predict bent's peak drift ratio (training data set)}
\label{table:Perm_Imp}
\begin{tabular}{lccccc}
\hline
Feature           & RR & SVR & RF & AdaBoost & NN \\ \hline
PGA               & 0.09                      & 0.1          & 0.06        & 0.05              & 0.07        \\
PGV               & 0.001                     & 0.005        & 0.001       & 0.002             & 0.005       \\
PGD               & 0.01                      & 0.04         & 0.01        & 0.02              & 0.004       \\
CAA               & 0.02                      & 0.09         & 0.04        & 0.02              & 0.04        \\
CAV               & 0.001                     & 0.002        & 0.002       & 0.001             & 0.0         \\
CAD               & 0.004                     & 0.005        & 0.001       & 0.001             & 0.0         \\
CSA               & 0.07                      & \textbf{0.14} & 0.08          & 0.05              & 0.13          \\
CSV               & 0.0                       & 0.002         & 0.001         & 0.0               & 0.001         \\
CSD               & 0.006                     & 0.007         & 0.002         & 0.005             & 0.001         \\
GMSa              & 0.11                      & 0.12          & 0.1           & 0.12              & 0.09          \\
{$Sa_1$}     & \textbf{0.15}             & \textbf{0.14} & \textbf{0.15} & \textbf{0.14}     & \textbf{0.16} \\ \hline
\end{tabular}
\end{table}

\subsection{Performance evaluation}

\begin{figure*}
	\centering
		\includegraphics[scale=0.90]{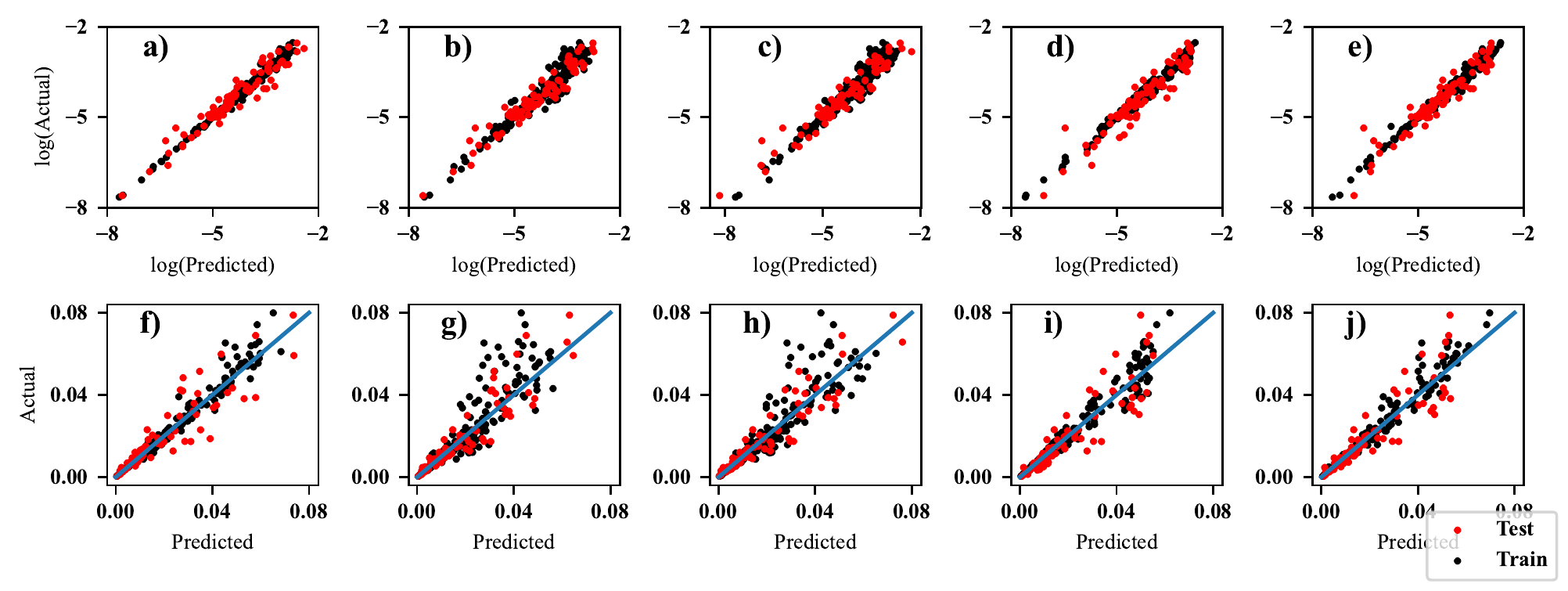}
	  \caption{Performance plots for PSDMs of SMA link strain developed via five different ML techniques: (a,f) neural network; (b,g) support vector regression; (c,h) ridge regression; (d,i) AdaBoost; (e,j) random forest }\label{PP_SMA}
\end{figure*}

In order to find the final PSDMs, for each ML method, a reasonable amount of hyperparameter tuning was performed until an efficient model was achieved. The values of two performance measures, i.e., $R^2$ and MSE, for all the final PSDMs of the bridge with SRR columns developed via various ML methods (on the test data set) are listed in Table~\ref{ModelQuality}. It is observed that, for almost all of the EDPs of interest, the neural network models have the best predictive performances (highest $R^2$ and lowest MSE) in comparison to other models. The only exception is for the low-cycle fatigue damage factor of the ED links, for which the AdaBoost model slightly outperforms the neural network model (with an $R^2$ of 0.8 vs. 0.78). The neural network models are found to only marginally outperform the support vector regression models. Except for the PSDMs developed for the bent's residual drift, max. SMA link strain, max. ED link strain, and max. ED link damage factor, the MSE values for all other neural network models are below 0.10. One should note that an MSE of 0.10 in the logarithmic space is equivalent to almost 10\% error in the real space. The max. SMA link strain is found to be predicted with much less error (lower MSE) through a neural network model compared to other ML models. The performance plots of the PSDMs developed for the max. SMA link strain using different ML techniques are compared in Fig.~\ref{PP_SMA}. As seen, all the models except the neural network model tend to slightly underestimate the response at higher levels. Given the foregoing, the PSDMs developed via the neural network method were used in the remainder of this study.


\begin {table*}[h!]
\centering
\caption{Model performance measures for different ML techniques}
\label{ModelQuality}
\begin{tabular}{lcccccccccccccc}
\hline
                              & \multicolumn{2}{c}{NN}        & \multicolumn{1}{l}{} & \multicolumn{2}{c}{SVR}       & \multicolumn{1}{l}{} & \multicolumn{2}{c}{Ridge} & \multicolumn{1}{l}{} & \multicolumn{2}{c}{AdaBoost}  & \multicolumn{1}{l}{} & \multicolumn{2}{c}{Random Forest} \\ \cline{2-3}  \cline{5-6} \cline{8-9} \cline{11-12} \cline{14-15}
Response                      & R2            & MSE           &                      & R2            & MSE           &                      & R2          & MSE         &                      & R2            & MSE           &                      & R2              & MSE             \\ \hline
Bent's peak drift ratio                  & \textbf{0.97} & \textbf{0.05} &                      & 0.95          & 0.09          &                      & 0.94        & 0.10        &                      & 0.96          & 0.06          &                      & 0.95            & 0.09            \\
Bent's residual drift ratio              & \textbf{0.84} & \textbf{0.11} &                      & \textbf{0.81} & \textbf{0.11} &                      & 0.78        & 0.12        &                      & 0.76          & 0.12          &                      & \textbf{0.81}   & \textbf{0.11}   \\
Max. cover concrete strain  & \textbf{0.93} & \textbf{0.09} & \textbf{}            & \textbf{0.94} & \textbf{0.09} &                      & 0.91        & 0.13        &                      & 0.92          & 0.10          &                      & 0.94            & 0.10            \\
Max. core concrete strain   & \textbf{0.94} & \textbf{0.09} &                      & \textbf{0.94} & \textbf{0.09} &                      & 0.92        & 0.10        &                      & 0.92          & 0.10          &                      & \textbf{0.94}   & \textbf{0.09}   \\
Max. longit. rebar strain  & \textbf{0.93} & \textbf{0.09} &                      & \textbf{0.93} & \textbf{0.10} &                      & 0.92        & 0.12        &                      & 0.92          & 0.11          &                      & 0.93            & 0.11            \\
Max. SMA link strain       & \textbf{0.94} & \textbf{0.11} &                      & 0.91          & 0.16          &                      & 0.89        & 0.18        &                      & 0.87          & 0.21          &                      & 0.89            & 0.18            \\
Max. ED link strain         & \textbf{0.94} & \textbf{0.12} &                      & \textbf{0.93} & \textbf{0.12} &                      & 0.92        & 0.13        &                      & 0.90          & 0.13          &                      & 0.92            & 0.13            \\
Max. ED link damage factor     & 0.78          & 0.31          &                      & 0.76          & 0.35          &                      & 0.72        & 0.41        &                      & \textbf{0.80} & \textbf{0.27} &                      & 0.78            & 0.32            \\
Max. bearing pad deformation                 & \textbf{0.95} & \textbf{0.05} &                      & \textbf{0.95} & \textbf{0.05} & \textbf{}            & 0.95        & 0.06        & \textbf{}            & 0.94          & 0.07          & \textbf{}            & \textbf{0.95}   & \textbf{0.05}   \\
Max. deck end's displacement               & \textbf{0.97} & \textbf{0.06} &                      & 0.96          & 0.07          &                      & 0.95        & 0.09        &                      & 0.94          & 0.11          &                      & 0.95            & 0.09            \\
Max. shear key deformation               & \textbf{0.94} & \textbf{0.05} &                      & \textbf{0.95} & \textbf{0.05} &                      & 0.94        & 0.06        &                      & 0.93          & 0.07          &                      & 0.93            & 0.07            \\ \hline
\end{tabular}
\end{table*}

\subsection{Interpretation}\label{interpret}
In this study, four model-agnostic methods are used to interpret some of the developed PSDMs - these methods include: (\emph{i}) Individual Conditional Expectation (ICE), (\emph{ii}) Partial Dependence Plot (PDP), (\emph{iii}) Accumulated Local Effects (ALE), and (\emph{iv}) Shapley Additive Explanations (SHAP). Among these methods, the first and the fourth ones are \emph{local} methods, i.e., they inspect the effects of input features on individual predictions. However, PDP and ALE are \emph{global} methods, i.e., they inspect the average effect of each input feature on the overall model prediction. Although SHAP is considered a local method, several global interpretation methods have been developed by aggregating SHAP for a large number of individual predictions~\citep{chen2021explaining}.
 
\begin{figure*}
	\centering
		\includegraphics[scale=0.90]{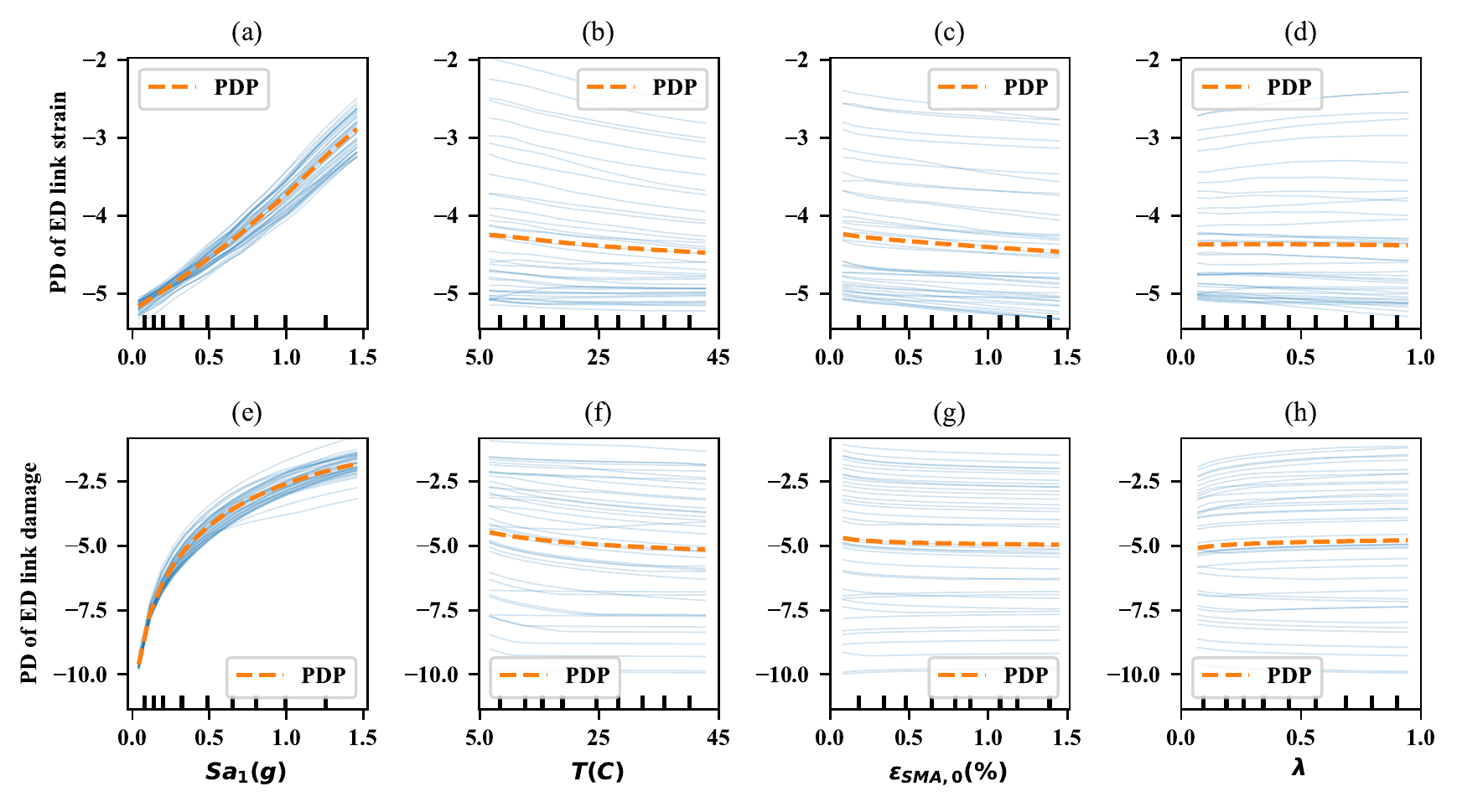}
	  \caption{ICE plots and PDPs for: (a-d) predicted max. ED link strain; (e-h) predicted max. ED link fatigue damage factor}\label{PD1}
\end{figure*}

\paragraph{ICE and PDP.}
ICE inspects the influence of a single feature on a single prediction by marginalizing the influences of the other features. Basically, an ICE plot shows the variation of a model's prediction with a selected feature, while the model's other input features are fixed at various values~\citep{goldstein2015peeking}. PDP is the global version of ICE, which is obtained by averaging many instances of ICE plots (e.g., see Fig.~\ref{PD1}). The main drawback of these methods is that, if some of the input features are correlated (which indeed is undesirable), then keeping one feature constant while varying the others to obtain the ICE plots (or subsequently, a PDP) could lead to misinterpretations.

The ICE plots (for 40 combinations of fixed feature values) and their corresponding PDPs investigating the effects of $Sa_1$, $T$, $\epsilon_{SMA,0}$, and $\lambda$ on the max. ED link strain and the ED link fatigue damage factor are shown in Fig.~\ref{PD1}. Note that there is no correlation between the selected input features, so ICE and PDP are usable here. As expected, the ground motion IM (i.e., $Sa_1$) is found to have the most significant effects on the selected EDPs - increasing $Sa_1$ results in the constant increase in the max. ED link strain (Fig.~\ref{PD1}(a)) and the max. ED link fatigue damage factor (Fig.~\ref{PD1}(e)). As seen in Fig.~\ref{PD1}(b), since the SMA links exhibit higher strength at higher temperatures (see Fig.~\ref{SMA_validation}(b)), thereby increasing the lateral strength of the SRR columns, the joint opening, and subsequently, the max. ED link strain decreases with $T$ (up to 5\% on average). For the same reason, higher $T$  is found to reduce the max. ED link fatigue damage factor by ~40\% on average (Fig.~\ref{PD1}(f)). According to Fig.~\ref{PD1}(c), increasing $\epsilon_{SMA,0}$ from 0 to 1.5\% decreases the ED link strain by ~5\% on average. This finding is justified by the fact that prestraining the SMA links increases their energy dissipation capacity~\citep{akbarnezhad2022seismic}, thereby reducing the bent's peak drift demands and joint opening. Due to their decreased strain demands, the max. fatigue damage factor of the ED links is also found to decrease with $\epsilon_{SMA,0}$ up to ~25\% (Fig.~\ref{PD1}(g)). Per Fig.~\ref{PD1}(d), decreasing $\lambda$ from 1 to 0 does not affect the ED link strain demands in an average sense. However, for the examples with higher ED link strain demands, decreasing $\lambda$ decreases the ED link strain by ~5\% because the energy dissipation of the SRR columns increases~\citep{akbarnezhad2022seismic}. Likewise, the average ED link damage factor decreases by ~25\% as $\lambda$ is reduced from 1 to 0 (Fig.~\ref{PD1}(h)). 

\paragraph{ALE.}
To tackle the inadequacy of the ICE and PDP methods in the presence of correlated features, one solution is to use conditional probability instead of marginal probability (for marginalizing the effects of the complementary set of features) to avoid the predictions associated with unrealistic combinations of correlated feature values. However, such an approach would let the correlated features influence the effect of the single feature of interest that is being investigated. To solve this issue, in the ALE approach, the gradients of the predictions with respect to the feature of interest are first computed, and then, these gradients are averaged over the conditional distribution of other features (local effects). Finally, a constant offset is applied to the computed value to make it centered - i.e., to make the mean effect zero. Integration of the local effects between two values of \textit{a} and \textit{b} is interpreted as the change in the prediction purely due to the feature's variation from \textit{a} to \textit{b}. This change in prediction can be seen as an independent measure of the effect of the feature on the prediction output~\citep{apley2020visualizing, mangalathu2022machine, molnar2020interpretable2}. In summary, the ALE value at each point represents the contribution of the feature of interest to the model prediction compared to the model's average prediction. It is noted that the values of the ALE and the PDPs are not expected to be the same, but their slopes can be interpreted similarly.

\begin{figure*}
	\centering
		\includegraphics[scale=0.65]{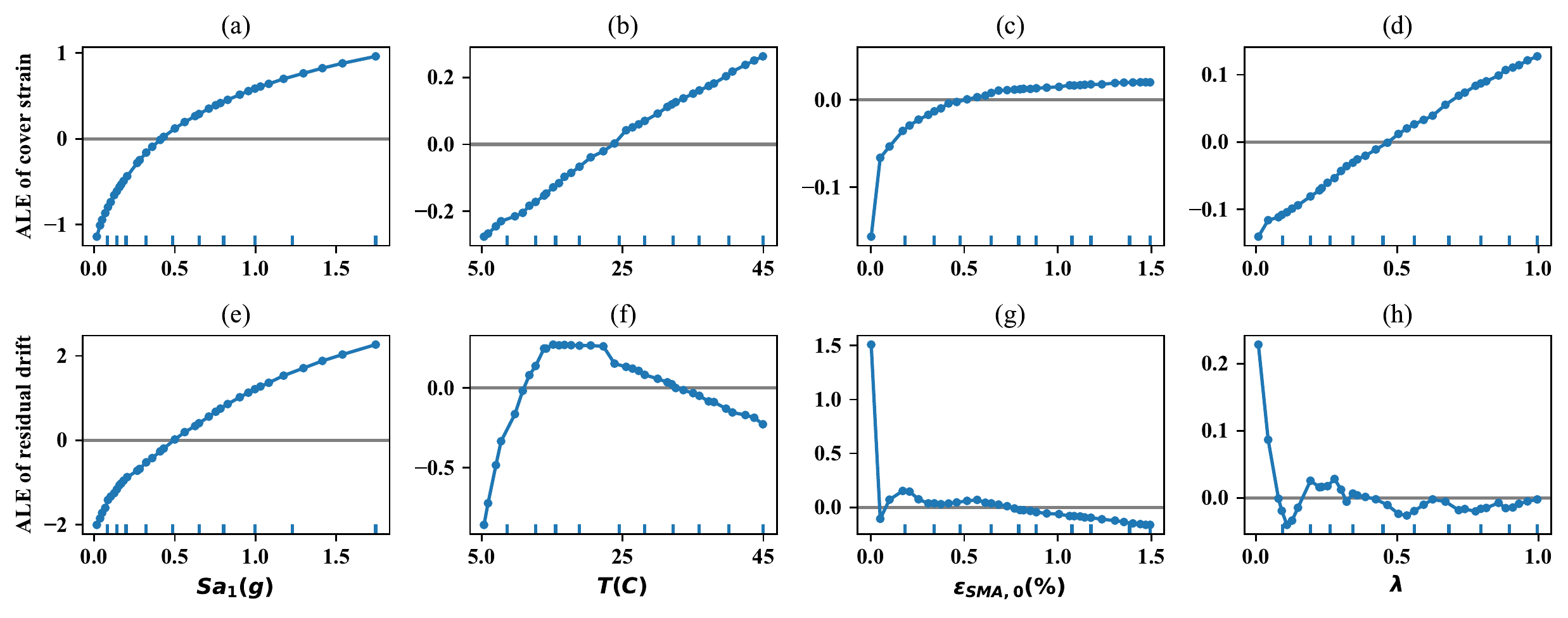}
	  \caption{ALEs of $Sa_1$, $T$, $\epsilon_{SMA,0}$, and $\lambda$ on: (a-d) predicted max. cover concrete strain; (e-h) predicted bent's residual drift}\label{ALE1}
\end{figure*}

Fig.~\ref{ALE1} shows the ALE plots aimed to explore the effects of the input features $Sa_1$, $T$, $\epsilon_{SMA,0}$, and $\lambda$ on the max. cover concrete strain and the bent's residual drift predictions. As expected, the positive slopes of the ALE plots in Fig.~\ref{ALE1}(a) and (e) verify that increasing $Sa_1$ would lead to increases in the max. cover concrete strain (due to increased curvature demands) and the bent's residual drift (due to increased plastic deformations under large displacement demands). According to the positive slope of the ALE plot in Fig.~\ref{ALE1}(b), the predicted max. cover concrete strain increases with the ambient temperature, too. This is because of the strength increase of the SMA links that would result in higher moment demands. The ALE plot for the effect of the ambient temperature on the bent's residual drift (Fig.~\ref{ALE1}(f)) exhibits positive and negative slopes at lower and higher ranges of $T$, i.e., a dual effect. Specifically, the initial SMA link strength increase induced by the temperature increase from 5$^\circ$C to about 20$^\circ$C results in higher residual drift demand predictions (likely because of the moment demand increase). However, further increase in the transformation stresses of the SMA links due to the temperature increase beyond 20$^\circ$C seem to actually have a decreasing effect on the bent's residual drift (likely because of the self-centering capacity increase). Per Fig.~\ref{ALE1}(c), increasing the initial strain of the SMA links, $\epsilon_{SMA,0}$, from 0 to 1.5\% is found to lead to higher max. cover concrete strain predictions (likely because of the moment demand increase). In terms of residual drift predictions, $\epsilon_{SMA,0}$ is found to have a slight decreasing effect (Fig.~\ref{ALE1}(g)), as higher prestraining could slightly increase the energy dissipation of SRR columns~\citep{akbarnezhad2022seismic}. According to Fig.~\ref{ALE1}(d), reducing self-centering coefficient, $\lambda$, from 1 to 0 (which results in higher energy dissipation, hence lower displacement demands) is observed to decrease the max. cover concrete strain predictions. However, $\lambda$ is found not to have a clear and significant effect on the bent's residual drift predictions (Fig.~\ref{ALE1}(h)).

\paragraph{SHAP.}
The SHAP method gives an insight into the effects of different input features on the model prediction through the \emph{Shapley/SHAP values}. The SHAP value is a solution concept in game theory to fairly distribute the payouts to players depending on their contributions to the total payout~\citep{shapley1953quota}. In the context of model interpretation, the SHAP value of each feature for a certain prediction quantifies that feature's contribution to pushing the prediction lower or higher from the model's average prediction or base value. Accordingly, the sum of the SHAP values of all features for a certain prediction is equal to the difference between that prediction and the base value. In that sense, one can describe the SHAP method as a local version of the ALE method to inspect individual predictions with respect to the average prediction. However, the SHAP values are computed entirely differently~\citep{vstrumbelj2014explaining, lundberg2017unified}. Though the SHAP method is a local interpretation method, aggregating the SHAP values for many prediction examples results in the global SHAP dependence plot, which is further discussed later on.

\begin{figure*}
	\centering
		\includegraphics[scale=0.95]{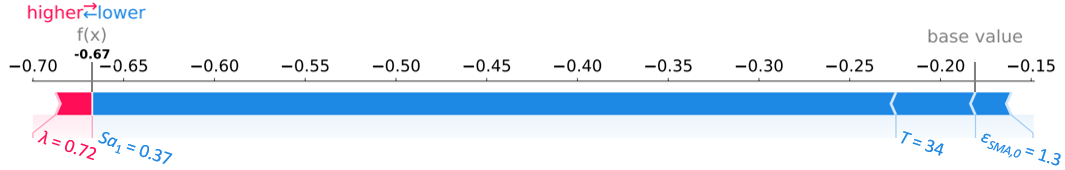}

	  \caption{Force plot for a single prediction of bent's peak drift}\label{shap1}
\end{figure*}

Fig.~\ref{shap1} illustrates the contributions of different features to the bent's peak drift ratio for a single prediction via a \emph{force plot}, which is a scaled visualization of the SHAP values. The feature values for the examined prediction are $Sa_1=0.37g$, $T=34^{\circ}$C, $\epsilon_{SMA,0}=1.3\%$, and $\lambda=0.72$. According to the force plot, the natural logarithm of the drift for this example is 0.49 smaller than the base value (-0.18). Moreover, it is observed that, for this certain prediction, $Sa_1$, $T$, and $\epsilon_{SMA,0}$ have pushed the natural logarithm of the peak drift ratio below its average by 0.45, 0.04, and 0.02, respectively. On the contrary, $\lambda$ has pushed the natural logarithm of the peak drift ratio above its average by 0.02.

\begin{figure}
	\centering
		\includegraphics[scale=0.50]{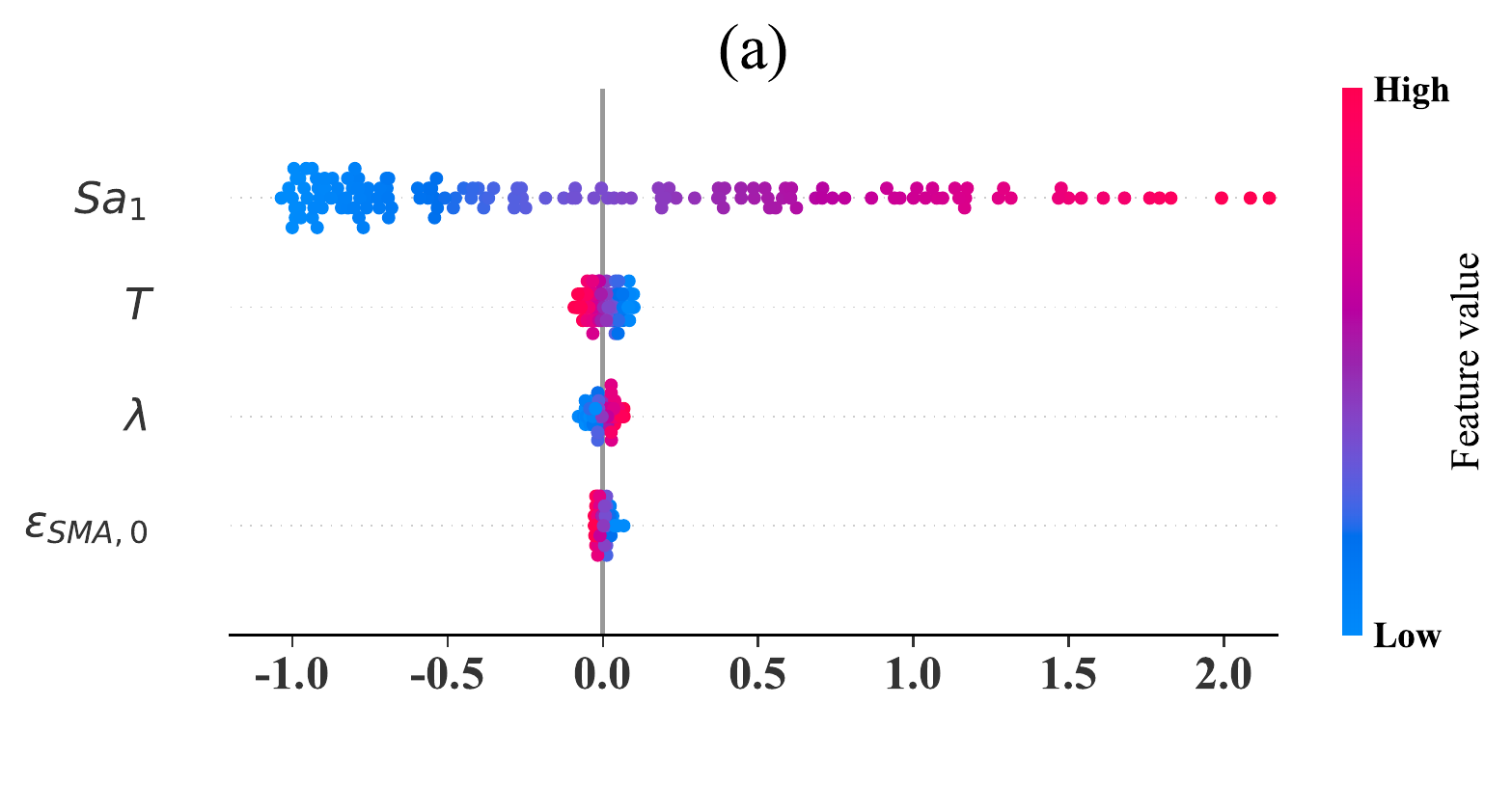}
		\includegraphics[scale=0.50]{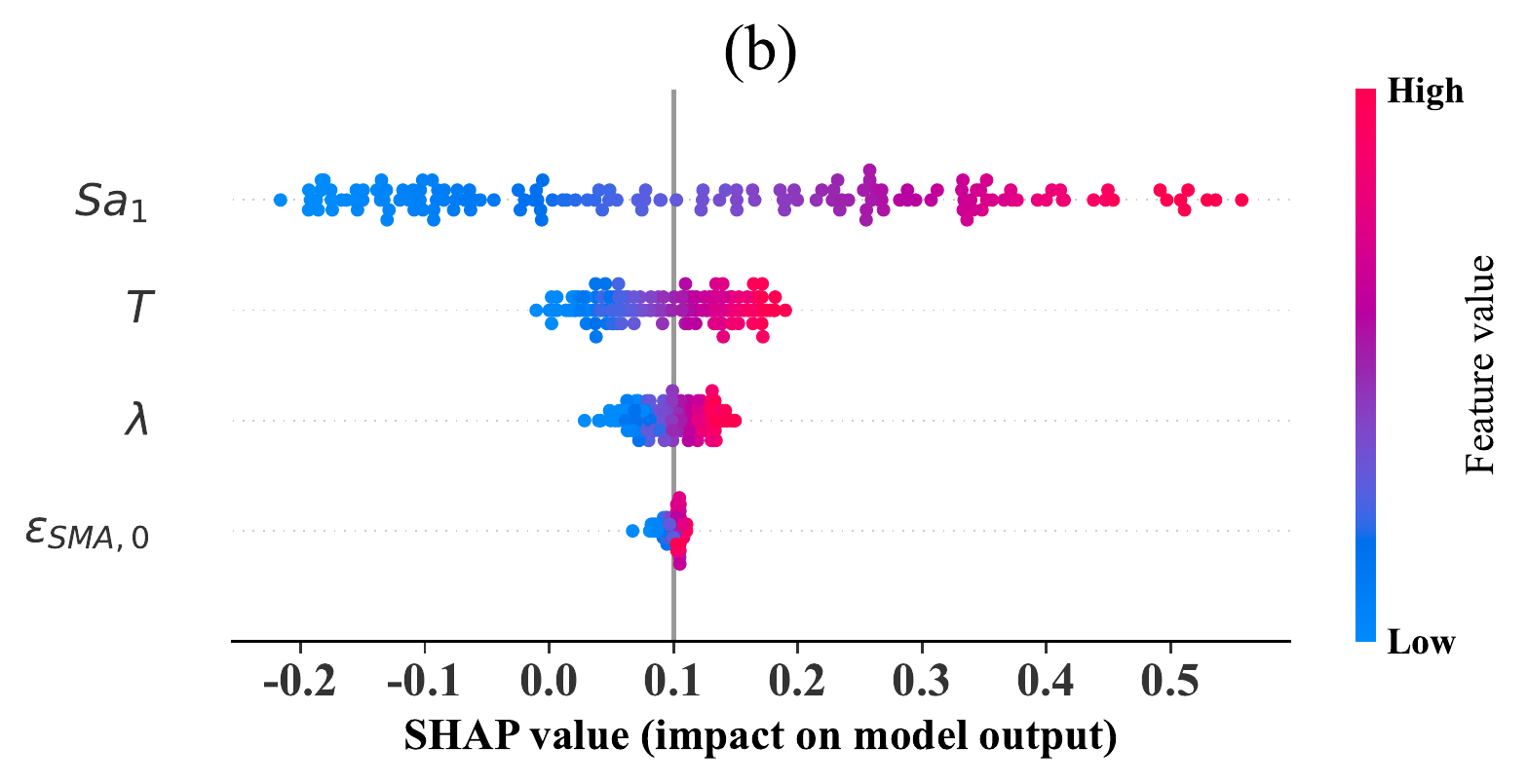}
	  \caption{Summary of SHAP values for: (a) peak bent drift ratio predictions; (b) max. longitudinal rebar strain predictions}\label{shap2}
\end{figure}

Fig.~\ref{shap2} presents the SHAP summary plots for the bent's peak drift and the max. longitudinal rebar strain PSDMs. Each summary plot summarizes the SHAP values for all the features and all the prediction instances within the data set - each dot represents the contribution of one feature to one prediction instance. The order of the features on the y-axis is determined based on their average importance, with the most important feature appearing at the top. Per Fig.~\ref{shap2}, $Sa_1$ has the most significant effects on the predictions of both EDPs and the higher $Sa_1$ is, the higher the selected EDP predictions are. As expected per earlier interpretations, the influences of $\epsilon_{SMA,0}$, $\lambda$, and $T$ on both EDPs are not as significant as that of $Sa_1$. As seen, $T$ is the second important feature for both prediction models. It is noted that since the SMA links exhibit higher strength at higher temperatures, thereby increasing the bent's lateral strength, its peak drift predictions decrease as $T$ increases (Fig.~\ref{shap2}(a)). However, the increased lateral strength leads to higher longitudinal rebar strain demands (Fig.~\ref{shap2}(b)). Per Fig.~\ref{shap2}(a), lower values of $\lambda$ are found to negatively contribute to the bent's peak drift predictions because the energy dissipation of the SRR columns increases as $\lambda$ decreases \citep{akbarnezhad2022seismic}. Likewise, lower values of $\lambda$ are found to less significantly contribute to the max. longitudinal rebar strain predictions (Fig.~\ref{shap2}(b)). Fig.~\ref{shap2} also shows that, among the input features, $\epsilon_{SMA,0}$ has a relatively inconsequential influence on both EDP predictions.  

\section{Seismic fragility analyses}\label{frag_anal}

The objectives of this section are to: (1) develop multi-parameter fragility functions for the bridge with SRR columns; (2) examine the effects of the key SRR column design parameters on the developed fragility functions while considering the effect of ambient temperature; and (3) compare the fragility functions of the bridge with SRR columns with those of the bridges of monolithic RC and PT columns.

\subsection{Fragility function development}\label{frag_func}
\begin{figure}
	\centering
		\includegraphics[scale=0.90]{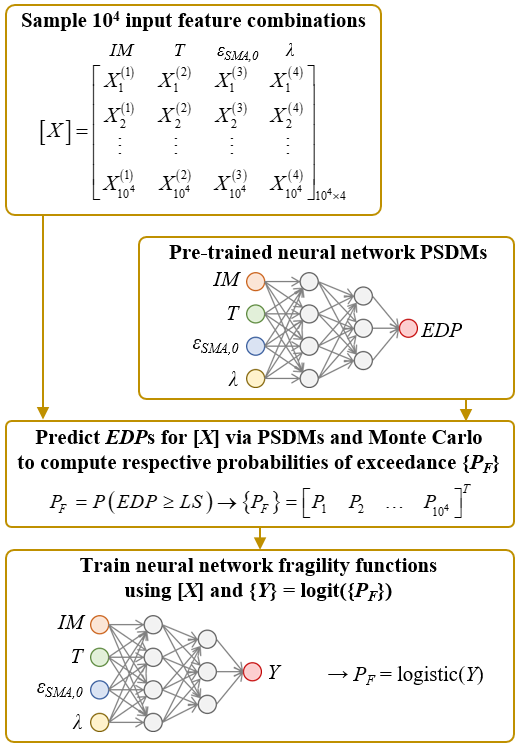}
	  \caption{Procedure of fragility function development}\label{flowchart}
\end{figure}

Fragility functions were developed for 11 DSs related to eight different bridge components, which are summarized in Table~\ref{Capacity models}. It is worth mentioning that, since the predicted maximum low-cycle fatigue damage factors for the ED links were rather small (rarely exceeding 0.2 for $Sa_1$ up to 1.5g), no meaningful damage states representing fatigue could be defined. A two-stage approach (see Section~\ref{ML_based}) was used to develop the fragility functions for the bridge with SRR columns. For this purpose, according to the procedure illustrated in Fig.~\ref{flowchart}, initially, a large number (in the order of $10^4$) of combinations of input feature values were generated using the Latin Hypercube sampling and based on their preselected distributions (see Table~\ref{table_input}). Subsequently, the PSDMs developed in Section~\ref{PSDM} were used to predict the EDPs of interest corresponding to each sampled combination of input features. The predicted EDPs were then used along with Monte Carlo simulations to compute the exceedance probabilities, $P_f$, of various limit states (LSs) according to their corresponding capacity models, which were assumed to be of log-normal distributions (Table~\ref{Capacity models}). Finally, neural network models with ReLU and linear activation functions at their intermediate and last layers, respectively, were trained to represent the fragility functions. In order to ensure that the exceedance probability predictions of the fragility functions are meaningful, i.e., between 0 and 1, the target values in the training data sets (exceedance probabilities) were initially transformed into a range of ($-\infty, +\infty$) using the \emph{logit} function. Therefore, the outputs of the trained models needed to be transformed back into a range of [0, 1] using the \emph{logistic} function.

\begin {table*}[h!]
\centering
\caption{Considered capacity distributions for different damage states}
\label{Capacity models}
\begin{tabular}{lccccccc}
\hline
Damage State ID & Description & EDP & $ln$(Median, $S_c$) & Dispersion, $\beta_C$ \\ 
\hline
I-1 & Column longitudinal rebar yielding & Max. tensile strain & $f_y/E$ & 0.35 \\
I-2 & Column longitudinal rebar fracture & Max. tensile strain & 0.15 & 0.35 \\
II-1 & Column cover concrete spalling & Max. compressive strain & 0.005 & 0.35 \\
II-2 & Column core concrete crushing & Max. compressive strain &  $\epsilon_{cu}$ & 0.35    \\
III & SMA link permanent deformation & Max. tensile strain  & $\epsilon_L+\sigma_{Mf}/E_M$ & 0.35 \\
IV & Bearing pad yielding & Peak deformation & 7.5 in. & 0.35 \\
V & Shear key failure & Peak deformation & 3.5 in. & 0.35 \\
VI & Abutment unseating & Peak deck end displacement  & 21 in. & 0.35 \\
VII & Bent major residual deformation & Bent residual drift & 1\% & 0.35  \\
\hline
\end{tabular}
\end{table*}

The neural network models representing the fragility functions were trained similarly to the PSDMs (see Section~\ref{PSDM}), aiming to minimize the cost function while tuning the hyperparameters to avoid overfitting based on MSE and $R^2$. The final MSE and $R^2$ of the fragility functions for all the damage states on the test data set were in the orders of 0.01 and 0.99-1.0, respectively, implying that the most significant sources of error for the fragility predictions within this framework were the PSDMs. For instance, the MSE and $R^2$ score on the test set for the fragility function of the abutment unseating were 0.01 and 1.0, respectively.

\begin{figure}
	\centering
		\includegraphics[scale=0.7]{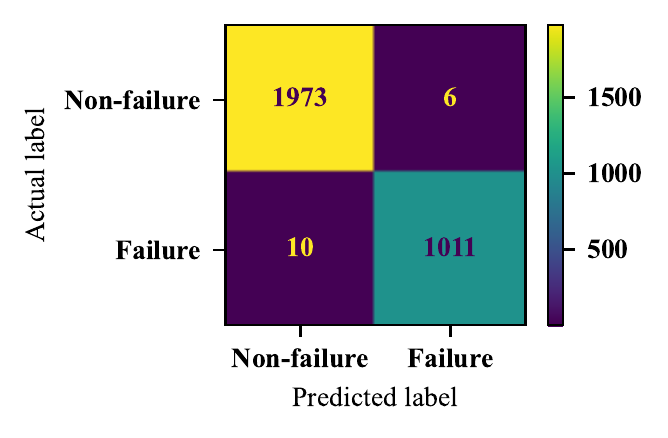}
	  \caption{Confusion matrix for  fragility prediction of bearings}\label{ConfM}
\end{figure}

To further evaluate the developed models based on the metrics used for classification models, e.g., recall and precision, the failure probabilities predicted for the test data were converted to non-failure/failure (0 or 1) binaries according to a threshold (e.g. 0.5). A higher precision indicates the model's higher power not to predict failure (1) for a real non-failure instance (0), while a higher recall indicates the model's higher power not to predict non-failure (0) for a real failure instance (1). All the developed fragility functions led to very high values of recall and precision (~1.0). For instance, the confusion matrix corresponding to the mapped fragility predictions for the test data (3000 data points) for bearing pads is shown in Fig~\ref{ConfM}. Accordingly, out of the 3000 examples, there are only 6 false positives and 10 false negatives, leading to recall and precision scores of about 0.99. 

\subsection{Effects of SRR column design parameters}
\begin{figure}[hb!]
    \centering
		\includegraphics[scale=0.83]{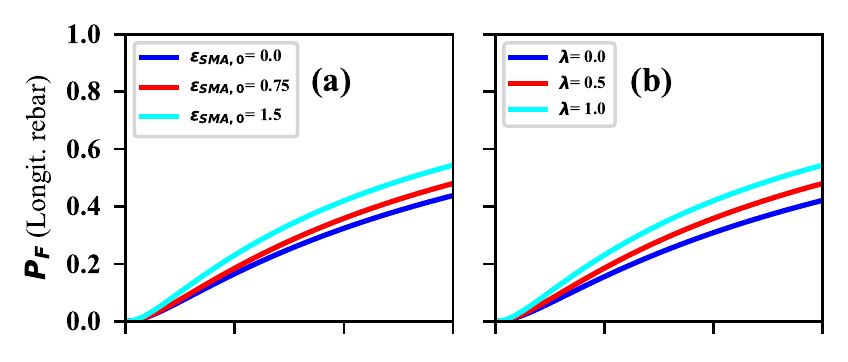}
		\includegraphics[scale=0.83]{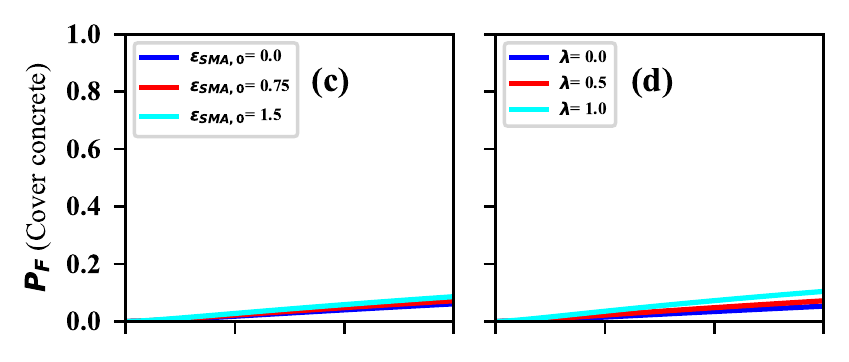}
		\includegraphics[scale=0.83]{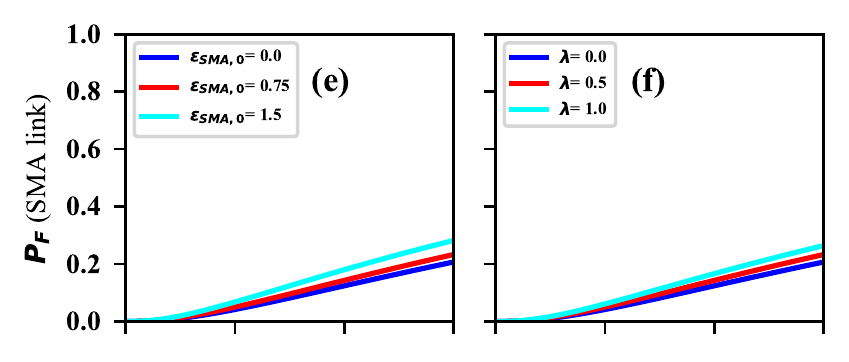}
		\includegraphics[scale=0.83]{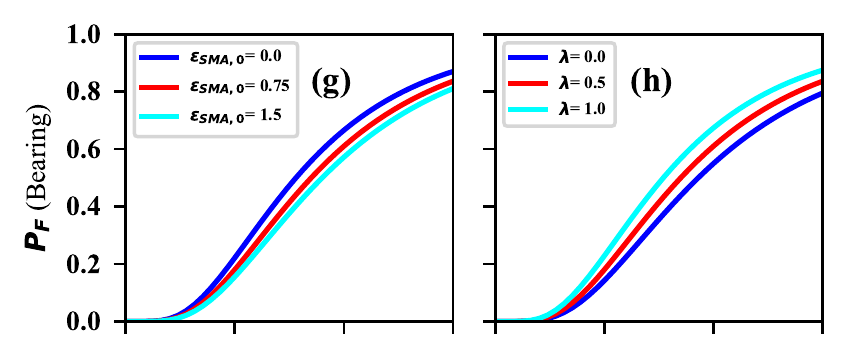}
		\includegraphics[scale=0.83]{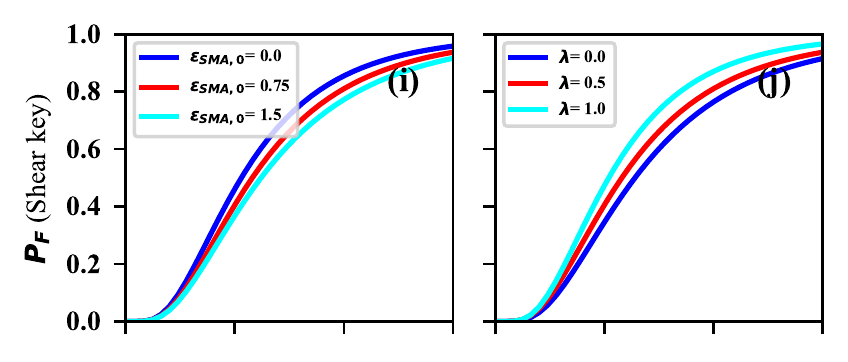}
		\includegraphics[scale=0.83]{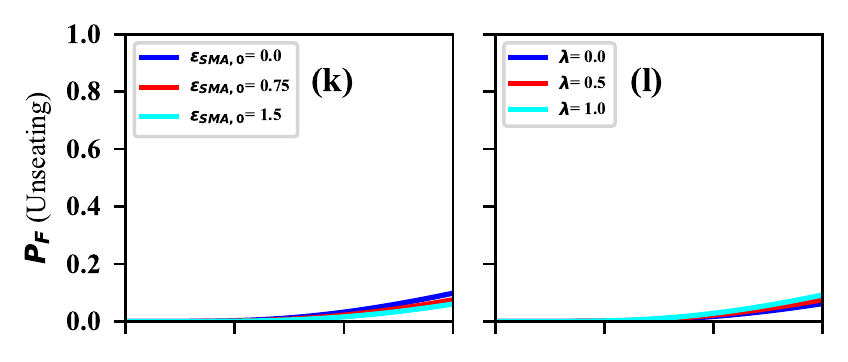}	
		\includegraphics[scale=0.83]{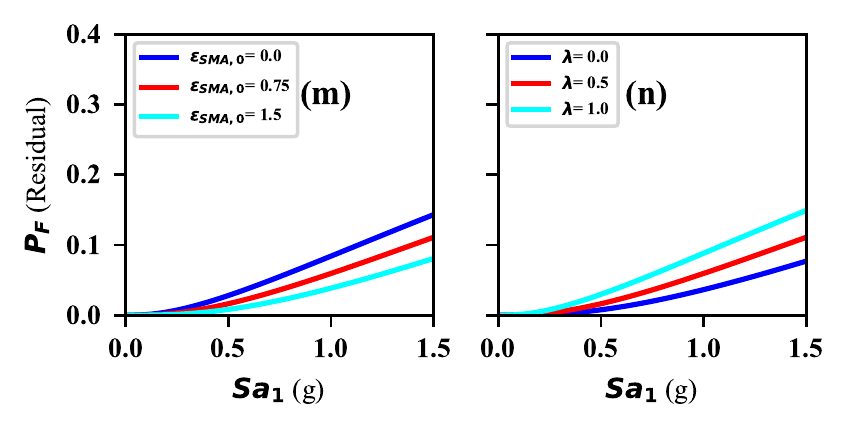}
	  \caption{Effects of $\epsilon_{SMA,0}$ and $\lambda$ on fragility curves of bridge with SRR columns}\label{FragSum1}
\end{figure}

The fragility curves obtained for seven DSs (I-1, II-1, III, IV, V, VI, and VII per Table~\ref{Capacity models}) and three different values of each of the SRR column key design parameters, i.e., $\epsilon_{SMA,0}$ and $\lambda$, are compared in Fig.~\ref{FragSum1}. For the fragility curves of various $\epsilon_{SMA,0}$ and $\lambda$ values, $\lambda$ and $\epsilon_{SMA,0}$ were fixed at their average values, i.e., 0.5 and 0.75\%, respectively. It is also worth mentioning that each curve was obtained considering the worst effect of ambient temperature.

According to Fig.~\ref{FragSum1}(g, i, k, m), the exceedance probabilities of damage states IV (bearing yielding), V (shear key failure), VI (unseating), and VII (major residual) decrease with increasing $\epsilon_{SMA,0}$. This is because such DSs are directly related to the bridge displacement demands, which are reduced as the energy dissipation of the SRR columns increases with increasing $\epsilon_{SMA,0}$ (see the discussion in Section~\ref{interpret}). Contrarily, it is observed in Fig.~\ref{FragSum1}(a, c) that the exceedance probabilities for damage states I-1 (rebar yielding) and II-1 (cover spalling) increase with $\epsilon_{SMA,0}$, which can be attributed to the increased base shear/moment demands for larger $\epsilon_{SMA,0}$. As expected, increasing the initial strains of the SMA links further leads to their increased strain demands, thereby increasing the exceedance probability for their permanent deformations (damage state IV, Fig.~\ref{FragSum1}(e)).

Before examining the effects of $\lambda$ on the obtained fragility curves, it is reminded that the $\lambda$ values 0.0 and 1.0 correspond to the SRR column designs with the maximum possible ED links to maintain self-centering and without any ED links, respectively~\citep{akbarnezhad2022seismic}. As seen in Fig.~\ref{FragSum1}(b, d, f, h, j, l, n), the exceedance probabilities of all of the examined DSs increase with $\lambda$. This is because, as discussed in Section~\ref{interpret}, increasing the amount of ED links (smaller $\lambda$) results in higher energy dissipation, and thus, smaller displacement demands. It is particularly interesting to observe that the probability of exceeding 1\% of bent residual drift (which is often considered large enough to necessitate bridge replacement~\citep{japan2002design, xiang2020probabilistic}), i.e., damage state VII, is also minimum when $\lambda = 0$, which indicates the importance of energy dissipation in reducing even the residual deformations. However, it is worth noting that, per ~\cite{akbarnezhad2022seismic}, the residual deformations of SRR columns are expected to increase as $\lambda$ gets smaller than 0.

\subsection{Comparison with other column designs}\label{col_frag_comp}
In order to assess the seismic vulnerability of the bridge with SRR columns relative to similar bridges but with monolithic RC and PT columns, the fragility functions of these bridges are compared in this section. The fragility functions of the latter bridges were developed through the same two-stage ML-based approach described earlier for the bridge with SRR columns.

\begin{figure}[hb!]
	\centering
		\includegraphics[scale=0.90]{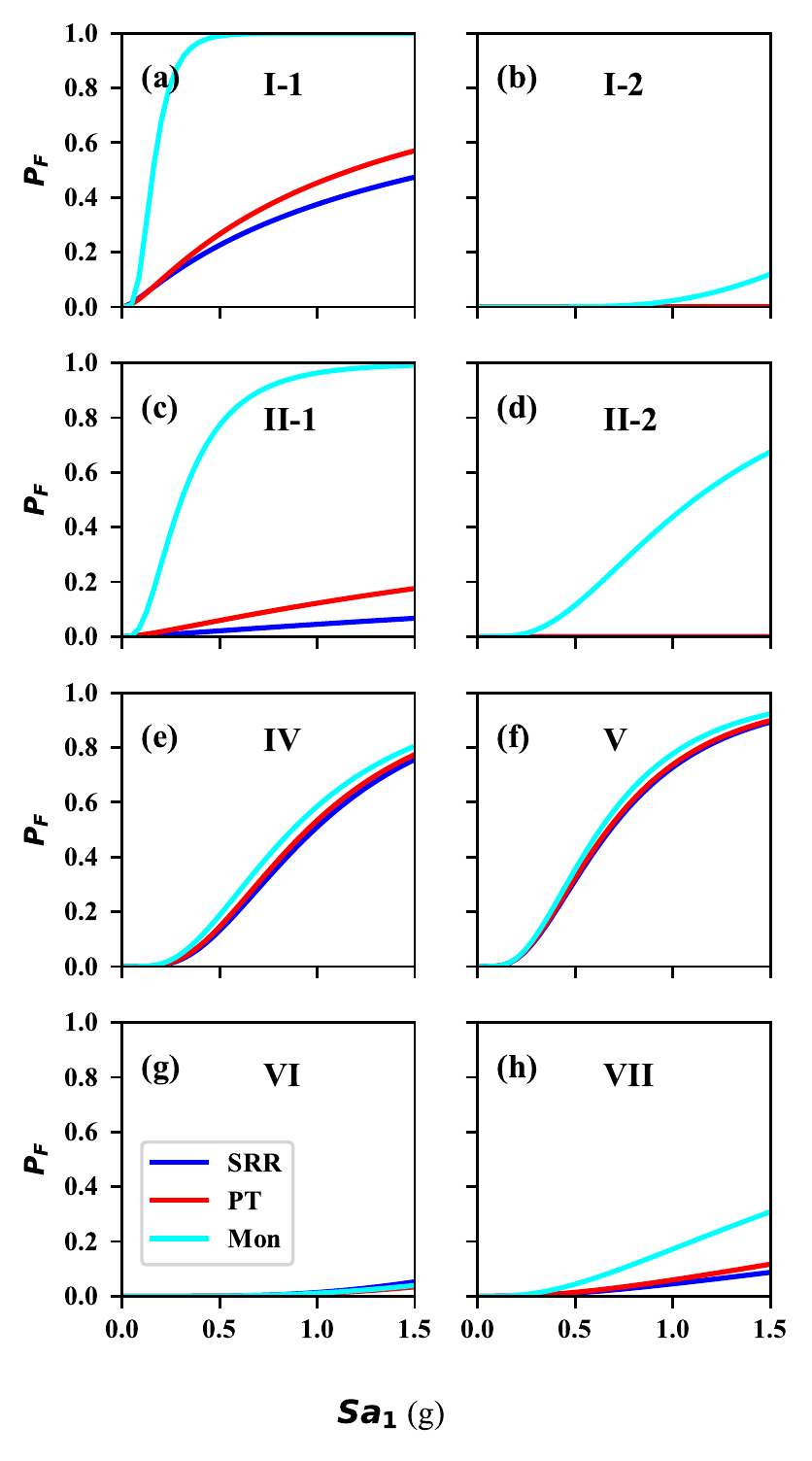}
	  \caption{Comparison of fragility curves for three different bridges: (a)  I-1; (b) I-2; (c) II-1; (d) II-2; (e) IV; (f) V; (g) VI; (h) VII 
	  }\label{frag}
\end{figure}

The fragility curves obtained for various DSs (I-1, I-2, II-1, II-2, IV, V, VI, and VII per Table~\ref{Capacity models}) of the three bridges are compared in Fig.~\ref{frag}. For the bridge of SRR columns, $\epsilon_{SMA,0}$ and $\lambda$ were fixed at 1.5\% and 0.0, while $T$ was assumed to be uncertain (see Table~\ref{table_input}). According to Fig.~\ref{frag}(a, b, c, d), the damage states associated with the columns' longitudinal rebar and concrete (I-1, I-2, II-1, and II-2) are all much likelier for the bridge of monolithic RC columns than for the bridges of PT and SRR columns. Specifically, the median $Sa_1$ at which column cover concrete spalling (damage state II-1) is predicted for the bridge of monolithic RC columns is almost 0.3g, while for the bridges of PT and SRR columns, the median $Sa_1$ for the same DS seem to be much larger than 1.5g (Fig.~\ref{frag}(c)). Likewise, though the predicted median $Sa_1$ for column core concrete crushing (damage state II-2) in the bridge of monolithic RC columns is almost 1.15g, this DS is not predicted to be exceeded in the bridges of PT and SRR columns (Fig.~\ref{frag}(d)). These findings primarily prove the damage mitigation efficacy of steel jacketing in the vicinity of the rocking joints. In terms of column longitudinal rebar yielding (damage state I-1), replacing the bridge's monolithic RC columns with PT and SRR columns is found to increase the respective median $Sa_1$ by more than 500\% (from ~0.2g to >1.25g; Fig.~\ref{frag}(a)). This was indeed expected because rocking joints inherently reduce the strain demands in the longitudinal rebar. Compared to rebar yielding, for the considered range of $Sa_1$, the probability of column longitudinal rebar fracture (damage state I-2) in all three bridges is extremely low (Fig.~\ref{frag}(b)).

Focusing on the fragility curves of the PT and SRR columns, it is observed that the SRR columns appear to be somewhat less seismically vulnerable than the PT columns of similar lateral load resistance and ED links (Fig.~\ref{frag}(a, c)). The lower probability of cover concrete spalling (damage state II-1) in the SRR columns compared to the PT columns (Fig.~\ref{frag}(c)) is mainly attributed to the lower axial force demands in the SRR columns than in the PT columns~\citep{akbarnezhad2022bridge}. However, The lower probability of longitudinal rebar yielding (damage state I-1) in the SRR columns compared to the PT columns (Fig.~\ref{frag}(a)) may be justified by the larger height of the steel jacket in the designed SRR columns compared to the designed PT columns.

Based on Fig.~\ref{frag}(e, f, g, h), as for other bridge damage states (IV, V, VI, and VII), the three bridges with different columns seem to be of similar fragility, except in terms of the bent's major residual deformation (damage state VII). Specifically, despite the lower energy dissipation capacity of rocking columns relative to the monolithic RC columns of similar lateral load resistance, which may lead to their higher drift ratio demands~\citep{lee2011performance}, the exceedance probabilities of bearing pad yielding (damage state IV), shear key failure (damage state V), and abutment unseating (damage state VI) at similar $Sa_1$ levels are very close (Fig.~\ref{frag}(e, f, g)). On the other hand, for the selected $Sa_1$ range, the probability of major residual drift (>1\%) in the monolithic RC bent is up to almost 3 times those in the bents of PT and SRR columns (Fig.~\ref{frag}(h)), demonstrating the self-centering efficacy of the rocking columns.


\section{Summary and conclusions}\label{conclusions}

This paper primarily aimed to evaluate the seismic fragility of a two-span RC highway bridge equipped with SRR columns through ML techniques. To this end, first, multi-parameter PSDMs were developed for multiple global and local EDPs using five different ML techniques, including (\emph{i}) ridge regression, (\emph{ii}) kernel support vector regression, (\emph{iii}) random forest, (\emph{iv}) adaptive boosting, and (\emph{v}) neural network. The input features of the initial PSDMs included a number of different ground motion characteristics as potential IMs, various material properties, two key SRR column design parameters (including the self-centering coefficient, $\lambda$, and the initial SMA link strain, $\epsilon_{SMA,0}$), and ambient temperature, $T$. To reduce the dimensionality of the initial PSDMs (without significant loss of their predictive power), the input features were subsequently reduced to only one “proficient” IM, $\lambda$, $\epsilon_{SMA,0}$, and $T$. In the next step, the performances of the PSDMs developed using various ML techniques were evaluated through $R^2$ and MSE metrics, and the most accurate PSDMs were selected. Following the selection of final PSDMs, the effects of $\lambda$, $\epsilon_{SMA,0}$, and $T$ on various EDPs were investigated through four model-agnostic local and global interpretation methods, namely, ICE, PDP, ALE, and SHAP. Subsequently, seismic fragility functions were developed for nine different DSs using neural networks, the selected PSDMs, and appropriate capacity models. Finally, the effects of $\lambda$ and $\epsilon_{SMA,0}$ on the fragility curves were investigated and the fragility curves of the bridge with SRR columns were compared with those of similar bridges with monolithic RC and PT columns. The major findings of this study are summarized below.

\begin{itemize}
\item Among the examined ground motion IMs, based on the permutation feature importance, the spectral acceleration at the period of 1 sec., $Sa_1$, was found to be more suitable to represent the ground motion intensity in the PSDMs of the bridge with SRR columns.
\item Overall, among the ML techniques used to develop the PSDMs for the bridge with SRR columns, the neural network was found to result in the most accurate prediction models, leading to $R^2$ scores as high as 0.97 and MSE as low as 0.05.
\item According to the PSDM interpretations, increasing $\epsilon_{SMA,0}$ decreases the SRR pier’s drift ratio demand, which in turn reduces the ED link strain demand. However, the prestressing of the SMA links slightly increases the moment demand in the SRR columns, leading to slightly higher longitudinal rebar and cover concrete strain demands. 
\item The PSDM interpretations also showed that decreasing $\lambda$, which amounts to increasing the ED links, slightly reduces the rebar and concrete damage in SRR columns. However, as long as $\lambda$ remains non-negative, the SRR bent’s residual drift ratio is not meaningfully affected by $\lambda$.
\item Based on the PSDM interpretations and because the SMA links exhibit higher strength at higher temperatures, increasing $T$ was found to slightly increase the longitudinal rebar and concrete strain demands. 
\item According to the fragility curves generated for the bridge with SRR columns, decreasing $\lambda$ to 0 (the minimum value allowed to ensure adequate self-centering) reduced the probabilities of exceedance of all the examined DSs. Contrarily, increasing $\epsilon_{SMA,0}$ to 1.5\% slightly reduced the probabilities of exceedance of displacement-dependent DSs (e.g., bearing damage and abutment unseating), while slightly increasing the probabilities of exceedance of strain-dependent DSs (e.g., longitudinal rebar yielding and concrete spalling).
\item Comparing the fragility curves for the bridges of SRR columns, monolithic RC columns, and PT columns showed that the SRR columns can seismically perform as well as (or even better than) the PT columns of similar lateral strengths. However, it was proven that both PT and SRR columns, which both benefit from rocking joints, can significantly outperform the monolithic RC columns in terms of longitudinal rebar and concrete damage, as well as residual deformations.
\end{itemize}

Overall, the findings of this study indicated that, even if the ambient temperature's uncertainty is taken into account, bridges of SRR columns can perform as well as (or even better than) bridges of PT columns. This conclusion besides the higher durability (due to the excellent corrosion resistance of NiTi) and easier repair (due to the easy replaceability of the SMA links) of SRR columns compared to PT columns can make SRR columns arguably suitable alternatives to PT columns.


%
%
%
\bibliographystyle{cas-model2-names}
%
\bibliography{cas-refs}
%
%

\end{document}